\documentclass{article}

\PassOptionsToPackage{sort, numbers, compress}{natbib}

\usepackage[final]{neurips_2020}

\makeatletter
  \def\NAT@spacechar{~}
\makeatother

\usepackage{amsmath}
\usepackage{amssymb}
\usepackage{amsthm}
\usepackage{booktabs}
\usepackage{chngcntr}
\usepackage{dsfont}
\usepackage{etoolbox}
\usepackage{graphicx}
\usepackage{hyperref}
\usepackage{microtype}
\usepackage{multirow}
\usepackage{paralist}
\usepackage{siunitx}
\usepackage{subcaption}
\usepackage{wrapfig}
\usepackage{xcolor}

\hypersetup{
  hyperfootnotes = false,
}

\definecolor{yale}{RGB}{0, 53, 107}

\hypersetup{
  colorlinks = true,
  linkcolor  = yale,
  urlcolor   = yale,
  citecolor  = yale
}

\usepackage{grfext}
\PrependGraphicsExtensions*{.png, .PNG}

\DeclareSIUnit\year{yr}

\newcommand{\comment}[1]{}

\usepackage{tikz}
\usepackage{pgfplots}

\usetikzlibrary{shapes}

\usepgfplotslibrary{colorbrewer}
\usepgfplotslibrary{groupplots}

\pgfplotsset{compat=1.16}

\DeclareMathOperator{\db}               {W_{\infty}}                           % Bottleneck distance
\DeclareMathOperator{\im}               {im}                                   % Image (of a function)
\DeclareMathOperator{\persistence}      {pers}                                 % Persistence (of a tuple)
\DeclareMathOperator{\probability}      {Pr}                                   % Probability
\DeclareMathOperator{\vertices}         {vert}                                 % Vertices
\DeclareMathOperator{\weight}           {w}                                    % Weight function

\newcommand{\betti}                  [1]{\ensuremath{\beta_{#1}}}              % Betti number
\newcommand{\cubicalcomplex}            {\mathrm{C}}                           % Cubical complex
\newcommand{\diagram}                   {\mathcal{D}}                          % Persistence diagram
\newcommand{\featurematrix}             {\mathbf{X}}                           % Feature matrix
\newcommand{\landau}[1]                 {\ensuremath{\mathcal{O}(#1)}}         % Landau symbols for complexity analysis
\newcommand{\naturals}                  {\mathds{N}}                           % Natural numbers
\newcommand{\reals}                     {\mathds{R}}                           % Real numbers
\newcommand{\simplicialcomplex}         {\mathrm{K}}                           % Simplicial complex
\newcommand{\timesteps}                 {\mathcal{T}}                          % Time steps
\newcommand{\volume}                    {\mathcal{V}}                          % Volumetric space

\newtheorem{theorem}{Theorem}

\title{Uncovering the Topology of Time-Varying fMRI Data using Cubical Persistence}

\author{
  Bastian Rieck\thanks{These authors contributed equally.}\\
  Dept. Biosystems~(D-BSSE)\\
  ETH Zurich \& Swiss Institute\\
  of Bioinformatics, Switzerland\\
  {\scriptsize\texttt{bastian.rieck@bsse.ethz.ch}}\\
  \And Tristan Yates\footnotemark[1]\\
  Dept. of Psychology\\
  Yale University \\
  New Haven, CT, USA \\
  {\scriptsize\texttt{tristan.yates@yale.edu}}\\
  \AND Christian Bock\\
  Dept. Biosystems~(D-BSSE)\\
  ETH Zurich \& Swiss Institute\\
  of Bioinformatics, Switzerland\\
  {\scriptsize\texttt{christian.bock@bsse.ethz.ch}}\\
  \And Karsten Borgwardt\\
  Dept. Biosystems~(D-BSSE)\\
  ETH Zurich \& Swiss Institute\\
  of Bioinformatics, Switzerland\\
  {\scriptsize\texttt{karsten.borgwardt@bsse.ethz.ch}}\\
  \And Guy Wolf\\
   Dept. of Math. and Stat. \\
   Univ. de Montr\'{e}al; Mila \\
   Montreal, QC, Canada \\
  {\scriptsize\texttt{guy.wolf@umontreal.ca}}\\
  \And Nicholas Turk-Browne\thanks{These authors jointly supervised this work; corresponding authors.}\\
   Dept. of Psychology\\
   Yale University \\
   New Haven, CT, USA \\
  {\scriptsize\texttt{nicholas.turk-browne@yale.edu}}\\
  \And Smita Krishnaswamy\footnotemark[2]\\
   Depts. of Gene.\ \& Comp.\ Sci.\ \\
   Yale University\\
   New Haven, CT, USA\\
  {\scriptsize\texttt{smita.krishnaswamy@yale.edu}}
}

\begin{document}

\maketitle

% \renewcommand*{\thefootnote}{\fnsymbol{footnote}}

% \footnotetext[1]{Department of Biosystems Science and Engineering, ETH Zurich, Switzerland}
% \footnotetext[2]{SIB Swiss Institute of Bioinformatics, Switzerland}
% \footnotetext[3]{Department of Psychology, Yale University, New Haven, CT, USA}
% \footnotetext[4]{Department of Mathematics and Statistics, Universit{\'e} de Montr\'{e}al; Mila, Montreal, QC, Canada}
% \footnotetext[5]{Department of Genetics, Yale University, New Haven, CT, USA}
% \footnotetext[6]{Department of Computer Science, Yale University, New Haven, CT, USA}

% \footnotetext[6]{These authors jointly supervised this work.}
% \footnotetext[2]{Department of Biosystems Science and Engineering, ETH Zurich, Switzerland}
% \footnotetext[3]{Department of Psychology, Yale University, New Haven, CT, USA}
% \footnotetext[4]{Department of Mathematics \& Statistics, Universit\'e de Montr\'eal, Montr\'eal, QC, Canada}
% \footnotetext[5]{Mila -- Quebec AI Institute, Montr\'eal, QC, Canada}

% Reset footnote counter since we are done with printing all the
% affiliations now.
\renewcommand*{\thefootnote}{\arabic{footnote}}
\setcounter{footnote}{0}

\begin{abstract}
  Functional magnetic resonance imaging~(fMRI)
  is a crucial technology for gaining insights into cognitive processes
  in humans. Data amassed from fMRI measurements result in volumetric
  data sets that vary over time. However, analysing such data presents
  a challenge due to the large degree of noise and person-to-person
  variation in how information is represented in the brain. To address
  this challenge, we present a novel topological approach that encodes
  each time point in an fMRI data set as a \emph{persistence diagram} of
  topological features, i.e.\ high-dimensional voids present in the
  data. This representation naturally does not rely on voxel-by-voxel
  correspondence and is robust to noise. We show that these
  time-varying persistence diagrams can be clustered to find meaningful
  groupings between participants, and that they are also useful in
  studying within-subject brain state trajectories of subjects
  performing a particular task. Here, we apply both clustering and
  trajectory analysis techniques to a group of participants watching the
  movie `Partly Cloudy'. We observe significant differences in both brain
  state trajectories and overall topological activity between adults and
  children watching the same movie.
\end{abstract}

%%%%%%%%%%%%%%%%%%%%%%%%%%%%%%%%%%%%%%%%%%%%%%%%%%%%%%%%%%%%%%%%%%%%%%%%
\section{Introduction}
%%%%%%%%%%%%%%%%%%%%%%%%%%%%%%%%%%%%%%%%%%%%%%%%%%%%%%%%%%%%%%%%%%%%%%%%

Human cognitive processes are commonly studied using functional magnetic
resonance imaging~(fMRI), amassing highly complex, well-structured, and
time-varying data sets across multiple individual subjects. fMRI uses
blood oxygen measurements of~3D brain volumes divided into
\emph{voxels}, i.e.\ 3D~pixels with dimensions in the \si{\milli\meter} range.
Voxels are measured over time while participants perform cognitive
tasks, resulting in time-varying activity measurements and an
activation function over the volume. The ultimate
goal of extracting higher-level abstractions from such data is primarily
impeded by two factors:
\begin{inparaenum}[(i)]
  \item the measurements are inherently noisy, due to changes in machine
    calibration, spurious patient movements, or environmental
    conditions, and
  \item there is a high degree of variability even between otherwise
    healthy brains~(e.g.\ in terms of the representation of stimulus and
    activity in the brain).
\end{inparaenum}
While these factors can be mitigated by certain experimental protocols
and pre-processing decisions, they cannot be fully eliminated.
This demonstrates the need for using representations that are to some
extent \emph{robust} with respect to noise and \emph{invariant} with
respect to isometric transformations in order to better capture
cognitively-relevant fMRI activity, particularly across populations
where anatomy--function relations may differ.

Traditional approaches largely ignore these factors, considering them
inevitable noise in the measurements. Voxel activity is often either
directly compared across different cognitive tasks, or the time-varying
activity of voxels in pre-defined brain regions sharing functional
properties is correlated to create a `functional connectivity' graph.
Our approach differs from existing approaches for fMRI data analysis in
two crucial ways, namely
\begin{inparaenum}[(i)]
  \item it is coordinate-free, providing a stable
  representation of high-level brain activity, even without
  a voxel-by-voxel match, and 
  \item it does \emph{not} require the creation of a correlation
    graph, or operate on any other approximated graph structure~(in
    contrast to the \textsc{Mapper} algorithm~\citep{Singh07}, for
    example).
\end{inparaenum}
Instead, our  method uses the `raw' voxel activations themselves as
a cubical complex, which we further characterise using time-varying
persistence diagrams that indicate the presences of topological
features such as voids of various dimensions in the voxel activations.
These topological features are naturally invariant to a variety of
shifts and noise~(see \autoref{sec:Methods} for more details). 
Our formulation enables the non-parametric analysis of fMRI data both
statically and dynamically, i.e.\ for assessing differences between
cohorts across time, and enabling insights into time-varying topological
brain state trajectories within cohorts or individuals. For individuals,
we calculate an averaged summary statistic over time that can be embedded
to explore population structure and variability statically, which we use
to organise subjects in our test set by age. Then, after partitioning
subjects into cohorts, we propose a novel method for producing
a time-varying trajectory of persistence diagrams that can be used to
quantify the progression and entropy of brain states. 
In summary, we make the following contributions:
{
  \begin{compactitem}
    \item We present a novel non-parametric framework for transforming
      time-varying fMRI data into time-varying topological representations.
    \item We empirically show that these representations
      \begin{inparaenum}[(i)]
      \item capture age-related differences, and
      \item shed light on the cognitive processes of age-stratified cohorts.
    \end{inparaenum}
    \item Finally, we show that our topological features are more
      informative for an age prediction task than other representations
      of the data set.
  \end{compactitem}
}

%%%%%%%%%%%%%%%%%%%%%%%%%%%%%%%%%%%%%%%%%%%%%%%%%%%%%%%%%%%%%%%%%%%%%%%%
\begin{figure}[t]
  \centering
  \subcaptionbox{\scriptsize fMRI images\label{sfig:fMRI stack}}{%
    \includegraphics[height=2.0cm]{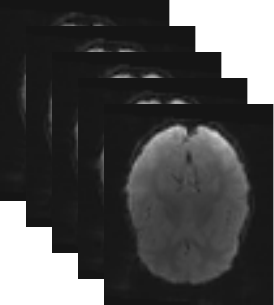}
  }%
  \subcaptionbox{\scriptsize fMRI volume\label{sfig:fMRI volume}}{%
    \includegraphics[height=2.0cm]{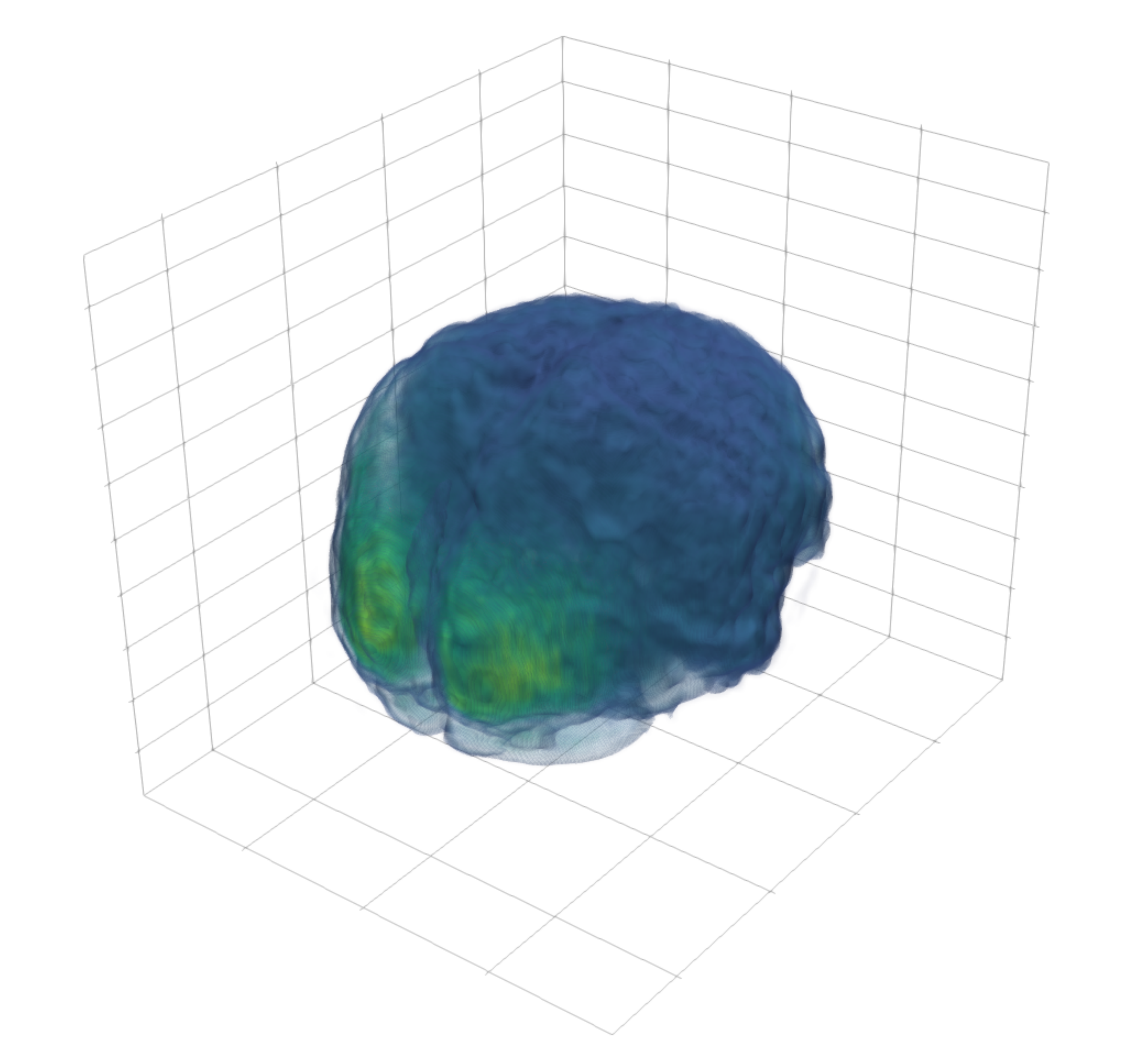}
  }%
  \subcaptionbox{\scriptsize Cubical complex\label{sfig:Cubical complex}}{%  	 
    \includegraphics[width=4.35cm]{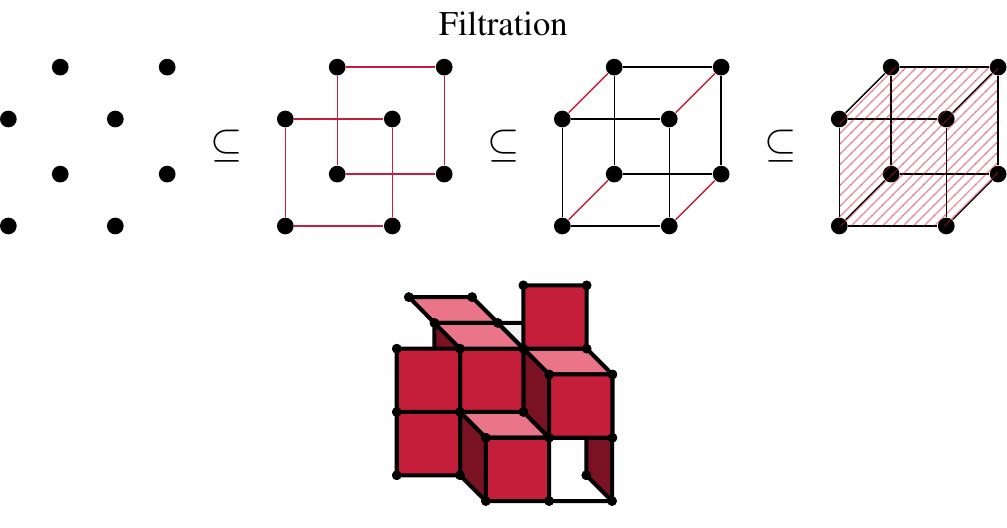}
  }
  \subcaptionbox{\scriptsize Persistence diagrams\label{sfig:Persistence diagrams}}{%
    \includegraphics[height=2.0cm]{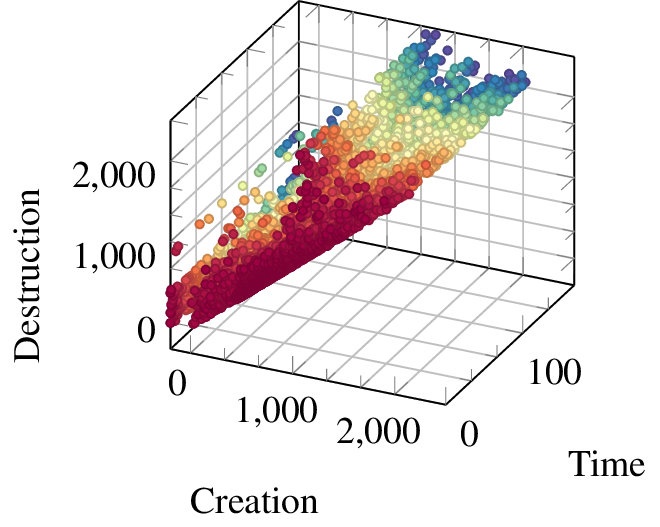}
  }
  \subcaptionbox{\scriptsize Persistence images\label{sfig:Persistence images}}{%
    \quad\includegraphics[height=2.0cm]{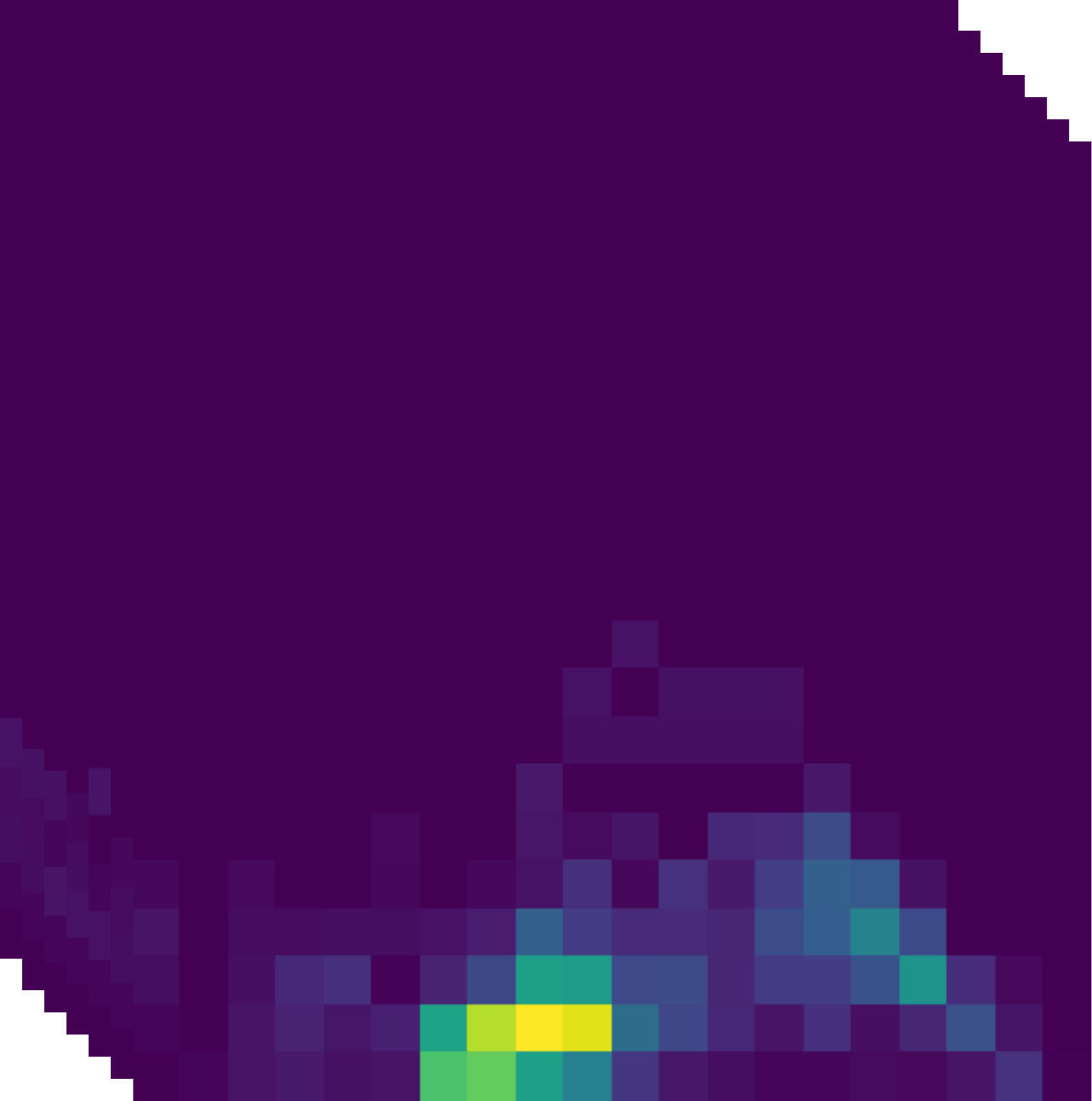}\quad
  }
  \caption{%
    A graphical overview of our method. We represent an fMRI
    stack~\subref{sfig:fMRI stack} as a volume~\subref{sfig:fMRI
    volume}, from which we create a sequence of cubical
    complexes~\subref{sfig:Cubical complex}. Calculating the persistent
    homology of this sequence results in a set of
    time-varying persistence diagrams~\subref{sfig:Persistence
    diagrams}; note that we only show the diagrams for a single
    dimension of the cubical complex.
    We calculate summary statistics from the diagrams~(not
    shown), and convert them to vectorial
    representations~\subref{sfig:Persistence images} for analysis
    tasks.
  }
\end{figure}
%%%%%%%%%%%%%%%%%%%%%%%%%%%%%%%%%%%%%%%%%%%%%%%%%%%%%%%%%%%%%%%%%%%%%%%%

%%%%%%%%%%%%%%%%%%%%%%%%%%%%%%%%%%%%%%%%%%%%%%%%%%%%%%%%%%%%%%%%%%%%%%%%
\section{Background on topological data analysis}\label{sec:Background}
%%%%%%%%%%%%%%%%%%%%%%%%%%%%%%%%%%%%%%%%%%%%%%%%%%%%%%%%%%%%%%%%%%%%%%%%

Topological data analysis~(TDA) recently started gaining traction in
machine learning~\citep{Carriere19, Hofer17, Hofer19, Hu19, Khrulkov18,
Kwitt15, Moor19, Ramamurthy19, Rieck19a, Rieck19b, Royer19, Zhao19}. TDA
is a rapidly-growing field that provides tools for analysing the shape
of data sets. This section provides a brief overview, aiming primarily
for intuition and less for depth~(see also \autoref{sec:Persistent
homology} for a worked example).
We refer to \citet{Edelsbrunner10} for
details. To our knowledge, this is the first time that TDA has been
\emph{directly} applied to fMRI data~(as opposed to applying it on
auxiliary representations such as functional connectivity networks).

%%%%%%%%%%%%%%%%%%%%%%%%%%%%%%%%%%%%%%%%%%%%%%%%%%%%%%%%%%%%%%%%%%%%%%%%
\paragraph{Simplicial homology.}
%%%%%%%%%%%%%%%%%%%%%%%%%%%%%%%%%%%%%%%%%%%%%%%%%%%%%%%%%%%%%%%%%%%%%%%%
%
The central object in algebraic topology is a simplicial
complex~$\simplicialcomplex$, i.e.\ a high-dimensional generalisation of
a graph, containing simplices of varying dimensions: vertices, edges,
triangles, tetrahedra, and their higher-dimensional counterparts.
A graph, for example, can be seen as a \mbox{1-dimensional} simplicial complex, containing vertices and edges.
Such complexes are primarily used to describe topological objects such as
manifolds\footnote{We will deviate from this notion later on in this paper but
follow the conventional exposition for now, which focuses primarily on
a simplicial view.}.
Simplicial homology refers to a framework for analysing the connectivity
of~$\simplicialcomplex$ via  matrix reduction algorithms,
assigning~$\simplicialcomplex$ a graded set of mathematical groups, the
homology groups.
Homology groups describe the topological features
of~$\simplicialcomplex$; in low dimensions~$d$, these features are
called \emph{connected components}~($d = 0$), \emph{tunnels}~($d = 1$),
and \emph{voids}~($d = 2$), respectively.
The number of \mbox{$d$-dimensional} topological features is referred to
as the $d$th Betti number~$\betti{d} \in \naturals$; it is used
to distinguish between different topological objects. For
example, a circle~(i.e.\ the \emph{boundary} of a disk) has Betti
numbers~$(1, 1)$ because there is a single connected component and
a single tunnel, while a filled square has Betti numbers~$(1,0)$.

%%%%%%%%%%%%%%%%%%%%%%%%%%%%%%%%%%%%%%%%%%%%%%%%%%%%%%%%%%%%%%%%%%%%%%%%
\paragraph{Persistent homology.}
%%%%%%%%%%%%%%%%%%%%%%%%%%%%%%%%%%%%%%%%%%%%%%%%%%%%%%%%%%%%%%%%%%%%%%%%
%
The analysis of real-world data sets, having no preferred scale at which
features occur, requires a different approach: Betti numbers cannot be
directly used here because they only represent counts, i.e.\ a single
scale. Endowing them with additional information leads to
\emph{persistent homology}, an extension of simplicial homology that
requires a simplicial complex~$\simplicialcomplex$ and an additional
function $f\colon\simplicialcomplex\to\reals$, such as an activation
function.
If $f$~only attains a finite set of function values $f_0 \leq f_1
\leq \dots \leq \dots f_{m-1} \leq f_m$, one can
sort~$\simplicialcomplex$ according to them, leading to
a \emph{filtration}---a nested sequence of simplicial complexes
\begin{equation}
  \emptyset = \simplicialcomplex_0 \subseteq \simplicialcomplex_1 \subseteq \dots \subseteq \simplicialcomplex_{m-1} \subseteq \simplicialcomplex_m = \simplicialcomplex,
  \label{eq:Filtration}
\end{equation}
with $\simplicialcomplex_i := \{\sigma \in K \mid f(\sigma) \leq f_i
\}$. Filtrations represent the evolution of~$\simplicialcomplex$
along~$f$.
Similar to the Watershed transform in image
processing~\citep{Roerdink00}, topological features can be
\emph{created}~(a new connected component might arise) or
\emph{destroyed}~(two connected components might merge into one) in a filtration.
Persistent homology efficiently tracks topological features across
a filtration, representing each one of them as a tuple $(f_i, f_j) \in
\reals^2$, with $i \leq j$ and $f_i, f_j \in \im(f)$.

%%%%%%%%%%%%%%%%%%%%%%%%%%%%%%%%%%%%%%%%%%%%%%%%%%%%%%%%%%%%%%%%%%%%%%%%
\paragraph{Persistence diagrams.}
%%%%%%%%%%%%%%%%%%%%%%%%%%%%%%%%%%%%%%%%%%%%%%%%%%%%%%%%%%%%%%%%%%%%%%%%
%
The tuples $(f_i, f_j)$ are collected according to their dimension~$d$
and stored in the $d$th \emph{persistence diagram}~$\diagram_d$, which
summarises all \mbox{$d$-dimensional} topological activity.
As a consequence of the calculation process, all points in~$\diagram_d$
are situated \emph{above} the diagonal.
The quantity $\persistence(x,y) := |y-x|$, i.e.\ the distance to the
diagonal~(up to a constant factor), of a point $(x, y) \in \diagram_d$
is called the \emph{persistence} of its corresponding topological
feature. Low-persistence features used to be considered `noise',
while high-persistence features are assumed to correspond to `real'
features of a data set~\citep{Edelsbrunner02}. Recent work cast some
doubts as to whether this assumption is justified~\citep{Bendich16};
in medical data, low persistence merely implies `low
reliability', \emph{not} necessarily `low importance.'

%%%%%%%%%%%%%%%%%%%%%%%%%%%%%%%%%%%%%%%%%%%%%%%%%%%%%%%%%%%%%%%%%%%%%%%%
\section{Related work}\label{sec:Related work}
%%%%%%%%%%%%%%%%%%%%%%%%%%%%%%%%%%%%%%%%%%%%%%%%%%%%%%%%%%%%%%%%%%%%%%%%

For fMRI analysis, the typical approach is to compare voxel activations
directly, but when one is interested in time-varying activity from
a continuous stimulus~(e.g.\ while watching a movie or resting),
voxel data is sometimes transformed into correlation matrices,
either calculated across \emph{time points}~\citep{Baldassano17}
or across \emph{voxels}~\citep{Wang15}. In the latter case, the goal is
to study functional connectivity, i.e.\ information about the
connectivity between brain regions sharing certain functional
properties.
Due to the size of the resulting matrices, one also often reduces the
dimensionality by applying an atlas parcellation~\citep{Schaefer17}.
Both of these representations are efficacious, with voxel-by-voxel
correlation matrices providing insights into the topology and dynamics
of human brain networks~\citep{Turk-Browne13}.
Moreover, for many multi-subject fMRI studies, \emph{shared response
models}~\citep{Chen15}, abbreviated as SRMs, have proven effective.
SRMs `learn' a mapping of multiple subjects into the same space,
enabling the detection of group differences, or the study of relations
between brain activity and movie annotations, for
example~\citep{Vodrahalli18}.
SRM was recently used to map voxel activity into a functional space~(as
opposed to an anatomical one) in order to study the brain representation of,
among others, visual and auditory information while receiving naturalistic
audiovisual stimuli~\citep{Kumar20}. Nevertheless, while it is one of
the most powerful techniques for extracting cognitively-relevant signals
from fMRI data, there is still room for improvement.

Previous work fusing (f)MRI analysis and topological data
analysis is either based on auxiliary~(topological)
representations~\citep{Saggar18, Sizemore19}, such as the \textsc{Mapper}
algorithm~\citep{Singh07} which operates on graphs, and requires numerous parameter choices,
or it makes use of functional connectivity
information~(information about connectivity between brain
regions sharing functional properties) and pre-defined regions of
interest~\citep{Anderson18, Chung13, Ellis19, Giusti16, Santos19}.
Some studies have investigated topological approaches on other
measuring modalities, such as structural MRI for anatomical
analyses~\citep{Chung09}, or diffusion MRI/DTI for studying white matter
integrity~\citep{Chung13}.
By contrast, our method operates \emph{directly} on fMRI volumes,
requiring neither additional location information nor auxiliary
representations. We will instead make use of \emph{cubical complexes},
for which we essentially replace triangles by squares and tetrahedra by
cubes~(see \autoref{sfig:Cubical complex} and the subsequent
section for details).
Cubical complexes and their homology are well-studied in algebraic
topology, but their use in real-world applications used to be limited to image
segmentation~\citep{Allili01}.
This changed with the rise of persistent homology, which was extended to
the cubical setting~\citep{Nanda12, Stroembom07, Wagner12}, leading to
cubical persistent homology~\citep{Dlotko18, Mrozek10, Wang16}.

%%%%%%%%%%%%%%%%%%%%%%%%%%%%%%%%%%%%%%%%%%%%%%%%%%%%%%%%%%%%%%%%%%%%%%%%
\section{A topology-based framework for fMRI data sets}\label{sec:Methods}
%%%%%%%%%%%%%%%%%%%%%%%%%%%%%%%%%%%%%%%%%%%%%%%%%%%%%%%%%%%%%%%%%%%%%%%%

In the following, we will be dealing with time-varying fMRI. By this, we
mean that we are observing an activation function~$f\colon\volume \times
\timesteps \to \reals$ over a~3D bounded volume~$\volume \subset \reals^3$
and a set of time steps~$\timesteps$. The alignment of~$\volume$ across
different subjects is highly non-trivial; we provide more details about
this at the beginning of \autoref{sec:Results}.
For $t \in \timesteps$, the function~$f(\cdot, t)$ is typically
visualised using either stacks of images~(\autoref{sfig:fMRI stack})
or volume rendering~(\autoref{sfig:fMRI volume}). While it would be
possible to analyse the topology of individual images~\citep{Bazin07},
we want a holistic view of the topology of~$\volume$.
To this end, we transform~$\volume$ into a \emph{cubical
complex}~$\cubicalcomplex$, i.e.\ an equivalent of a simplicial complex,
in which triangles and tetrahedra are replaced by
squares and cubes~(see \autoref{sfig:Cubical complex}).
Cubical complexes are perfectly suited to represent an fMRI
volume~$\volume$ because each voxel corresponds precisely to one cubical
simplex~(whereas if we were to use a simplicial complex, we would have
to employ interpolation schemes as there is no natural mapping from
voxels to tetrahedra; see \autoref{fig:Simplicial complex interpolation}
for more details).

%%%%%%%%%%%%%%%%%%%%%%%%%%%%%%%%%%%%%%%%%%%%%%%%%%%%%%%%%%%%%%%%%%%%%%%%
\paragraph{Terminology.}
%%%%%%%%%%%%%%%%%%%%%%%%%%%%%%%%%%%%%%%%%%%%%%%%%%%%%%%%%%%%%%%%%%%%%%%%
%
We assume that we are given a data set of~$n$ volumes
$\volume_1, \dots, \volume_n$, corresponding to~$n$
different individuals, and a set of~$m$ time steps
$\timesteps = \{t_1, \dots, t_m\} \subset \naturals$.
We use $\vertices(\volume_i)$ to denote the vertex~(i.e.\ voxel) set
of~$\volume_i$, and $f_i$ to denote its activation function, i.e.\
$f_i\colon\volume_i\times\timesteps\to\reals$,
Here, the activation functions are \emph{aligned} with respect to their
time steps; this is an assumption that greatly simplifies all subsequent
analysis steps. It does not impose a large restriction in practice.

\paragraph{Topological features from fMRI data.} We obtain topological features of each~$f_i$ following a three-step
procedure, namely
\begin{inparaenum}[(1)]
  \item cubical complex conversion,
  \item filtration calculation, and
  \item persistence diagram calculation.
\end{inparaenum}
The \emph{conversion} of a volume~$\volume_i$ to a cubical
complex~$\cubicalcomplex_i$ is simple, as~$\volume_i$ and
$\cubicalcomplex_i$ share the same cubical elements and connectivities.
Thus, the vertices of $\cubicalcomplex_i$ are the voxels of
$\volume_i$ and there are edges between neighbouring vertices as defined
by a regular 3D~grid, in which each vertex has six neighbours~(two per
coordinate axis).
These neighbourhoods implicitly  define the connectivity of higher-dimensional elements~(squares and
cubes).
We will use $\sigma$ to denote an element of a cubical
complex\footnote{These elements are the `simplices' of the cubical
complex, but we refrain from re-using the term `simplex' so as not
to confuse ourselves or the reader.}.
Next, we impose a \emph{filtration}---an ordering---of the
elements of~$\cubicalcomplex_i$. Since we want to analyse
topological features over time, we have to calculate one filtration
for every time step. Given $t_j \in \timesteps$, we assign the values
of~$f_i(\cdot, t_j)$ to~$\cubicalcomplex_i$.
We use the most natural assignment: each vertex~(voxel)
of~$\cubicalcomplex_i$ receives its activation value at time $t_j$,
while a higher-dimensional element~$\sigma$ is assigned a value recursively
via $f_i(\sigma, t_j) := \max_{v \in \vertices(\sigma)} f_i(v, t_j)$. We
then sort the cubical complex~$\cubicalcomplex_i$
in ascending order according to these values; in case of a tie,
a lower-dimensional element~(e.g.\ an edge) precedes a higher-dimensional
one~(e.g.\ a square).
Having obtained a filtration according to
\autoref{eq:Filtration}, we may now calculate the persistent
homology of~$\cubicalcomplex_i$ at time step~$t_j$, resulting
in a collection of \emph{persistence diagrams}. Since each~$\volume_i$ is
three-dimensional, we obtain a triple
$\left(\diagram_0^{(i, j)}, \diagram_1^{(i, j)}, \diagram_2^{(i, j)}\right)$
for every time step~$t_j$; persistence diagrams for~$d \geq 3$ are all
empty.
Notice that the calculation of persistence diagrams for
a participant~$i$ and a time step~$t_j$ can be easily parallelised
since we treat time steps independently.  Subsequently, we will
use $\diagram^{(i)}$ to denote the set of all persistence diagrams
associated with the $i$th participant.
We can plot the resulting persistence diagrams of each participant as
a set of diagrams in $\reals^3$, with the additional axis being used to
represent \emph{time}~(\autoref{sfig:Persistence diagrams}).

The filtration that we employ here is also known as a \emph{sublevel set
filtration}. Other filtrations~\citep{Anai20} could also be used~(our method is not
restricted to any specific one), but a symmetry
theorem~\citep{Cohen-Steiner09} states that unless we are willing to
modify the activation function values themselves we are not gaining any
more information about the topology of our input data. 
For the subsequent analyses, we will be dealing with collections of
persistence diagrams~$\diagram^{(i)}$. The space of persistence diagrams
affords several metrics~\citep{Cohen-Steiner07, Cohen-Steiner10}, but
they are computationally expensive and infeasible for the cardinalities
we are dealing with~(a typical persistence diagram of a participant
contains about \num{10000} features). We will thus be working with \emph{topological
summary statistics} and \emph{persistence diagram vectorisations}. 

\paragraph{Properties.}
Prior to delving deeper into our pipeline, we describe
some properties of our approach and why topological features are
advantageous. Topology is inherently coordinate-free, meaning that all
the features we describe are invariant to homeomorphism, i.e.\ stretching and bending.
Moreover, the persistence diagrams of spaces of
different cardinalities and scales can be compared, making it possible
to `mix' participants from studies with different imaging
modalities or resolutions~(of course, this should not be done
indiscriminately). Arguably the largest advantage of persistent homology
is its stability with respect to perturbations~\citep{Cohen-Steiner07,
Cohen-Steiner10, Skraba20}. This is quantified by
the following theorem, whose proof we defer to \autoref{sec:Stability
proof}.
\begin{theorem}
  Let~$f\colon\volume\to\reals$ and~$g\colon\volume\to\reals$ be two
  activation functions. Then their corresponding persistence
  diagrams~$\diagram_f$ and~$\diagram_g$ satisfy
    $\db(\diagram_f, \diagram_g) \leq \|f - g\|_{\infty}$, 
  where $\db$ denotes the bottleneck distance between persistence
  diagrams, defined as
  $\db(\diagram_f, \diagram_g) := \inf_{\eta\colon \diagram \to
  \diagram_g}\sup_{x\in{}\diagram_f}\|x-\eta(x)\|_\infty$, with
  $\eta\colon\diagram_f\to\diagram_g$ denoting a bijection between the
  points of the two diagrams, and $\|\cdot\|_\infty$ referring to the
  $\mathrm{L}_\infty$ norm.
\end{theorem}
The consequence of this stability theorem is that the persistence
diagrams that we calculate are stable with respect to perturbations,
provided those perturbations are of small amplitudes. This is
a desirable characteristic for a feature descriptor because it provides
us with well-defined bounds for its behaviour under noise. A more
precise version of this stability theorem
exists~\citep{Cohen-Steiner10}, but requires a more involved
setup\footnote{%
  We will have to show Lipschitz continuity for the functions, plus
certain other properties of the space~$\volume$.
}, which we leave for future work.
In general, we note that time-varying TDA is still a rather nascent
sub-field of TDA. A standard approach, namely the calculation of
`persistence vineyards'~\citep{Cohen-Steiner06}, resulting in
a decomposition of a time-varying persistence diagram into individual
`vines', is not applicable here because the changes between different
time steps are not infinitesimal~(there \emph{is} a large
amount of temporal coherence between consecutive time steps, but there
is no guarantee that the change between them is upper-bounded). It is
still unknown in our setting whether a vineyard representation with
unique vines exists at all~\citep{Munch13}.
We therefore prefer to treat the individual time steps as independent
calculations but note that future work should address a more efficient
computation by exploiting similarities between consecutive time steps.

%%%%%%%%%%%%%%%%%%%%%%%%%%%%%%%%%%%%%%%%%%%%%%%%%%%%%%%%%%%%%%%%%%%%%%%%
\paragraph{Implementation and complexity.}
%%%%%%%%%%%%%%%%%%%%%%%%%%%%%%%%%%%%%%%%%%%%%%%%%%%%%%%%%%%%%%%%%%%%%%%%
%
While the general calculation of persistent homology on high-dimensional
simplicial complexes is still computationally expensive, there are
highly-efficient algorithms for lower-dimensional
calculations~\citep{Bauer14a}.
Here, we use \texttt{DIPHA}~\citep{Bauer14b}, a distributed
implementation of persistent homology, as it implements an efficient
algorithm for computing topological features of cubical
complexes~\citep{Wagner12}.
Space and time complexity is linear in the number of voxels, so our
conversion process does not change the complexity of processing the
data. The persistent homology calculation has a time complexity of
$\landau{|\volume|}^{\omega}$, with $\omega
\approx 2.376$~\citep{Milosavljevic11}.
The distributed implementation of \texttt{DIPHA} is reported~\citep{Bauer14b}
to be capable of calculating persistent homology for $|\volume| \approx 10^9$,
making our pipeline feasible and scalable.
For the persistence image calculation in \autoref{sec:Dynamic analysis},
we use \texttt{Scikit-TDA}~\citep{scikit-tda}.
We make our code publicly available\footnote{\url{https://github.com/BorgwardtLab/fMRI_Cubical_Persistence}} to ensure reproducibility.

%%%%%%%%%%%%%%%%%%%%%%%%%%%%%%%%%%%%%%%%%%%%%%%%%%%%%%%%%%%%%%%%%%%%%%%%
\section{Results}\label{sec:Results}
%%%%%%%%%%%%%%%%%%%%%%%%%%%%%%%%%%%%%%%%%%%%%%%%%%%%%%%%%%%%%%%%%%%%%%%%

We evaluate our topological pipeline using open-source fMRI
data~\citep{richardson_development_2018}, available on the OpenNeuro
database~(accession number \texttt{ds000228}). The participants
comprised 33 adults~(18--39 years old; M~=~24.8, SD~=~5.3; 20~female) and 122
children~(3.5--12 years old; M~=~6.7, SD~=~2.3; 64~female) who watched the same
animated movie `Partly Cloudy'~\citep{partly_movie} while undergoing
fMRI. Please refer to \autoref{sec:fMRI pre-processing} and
\citet{yates_emergence_2020} for a full description of the
pre-processing.
The relevant outputs of these pre-processing steps are:
a $4$-dimensional~($x \times y \times z \times t$, with $x, y, z$ being
coordinates, and $t$ representing time) fMRI time series and
a whole-brain mask~(BM) for each individual subject.
The 4D volume of each participant has dimensions $65 \times 77 \times 60
\times 168$. Each of 168 time steps of the fMRI time series comprises 
\SI{2}{\second} of the movie and corresponds to the same point in the
movie for each subject; since for the first five time steps only a blank
screen was shown, we remove these plus two time steps to
account for the fMRI hemodynamic lag for all analyses.
We supplemented the whole-brain mask by also creating an
`occipital-temporal' mask~(OM) for each subject. This entailed finding the
intersection between an individual subject's
whole-brain mask and occipital, temporal, and precuneus regions of
interest defined from the Harvard--Oxford cortical atlas. If our
results reflect patterns relevant to cognitive processing, we would
expect similar---if not better---results using this occipital-temporal mask,
since it contains the regions most consistently involved in
movie-watching~(e.g.\ visual regions).
Last, we also calculated the `logical XOR' between the whole-brain mask
and the occipital-temporal mask; this mask~(XM) makes it possible to
study the relevance of topological features with respect to non-visual
regions~(including the frontal lobe) in the brain.
To prevent analysis bias, data were initially fully unlabelled during
the development of our pipeline. Later on, participants were assigned to
cohorts based on their age group, using the same bins as
\citet{yates_emergence_2020}; we initially did not know whether cohorts
were sorted in ascending or descending order. The actual ages were only
used in the age prediction experiment, which was performed \emph{after}
method development had ceased.

%%%%%%%%%%%%%%%%%%%%%%%%%%%%%%%%%%%%%%%%%%%%%%%%%%%%%%%%%%%%%%%%%%%%%%%%
\subsection{Static analysis based on summary statistics}\label{sec:Static analysis}
%%%%%%%%%%%%%%%%%%%%%%%%%%%%%%%%%%%%%%%%%%%%%%%%%%%%%%%%%%%%%%%%%%%%%%%%

%%%%%%%%%%%%%%%%%%%%%%%%%%%%%%%%%%%%%%%%%%%%%%%%%%%%%%%%%%%%%%%%%%%%%%%%
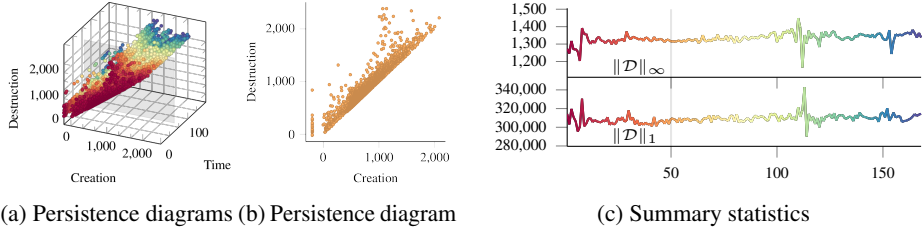
\begin{figure}[t]
  \centering
  \subcaptionbox{Persistence diagrams\label{sfig:Persistence diagram 3D}}{
    \includegraphics[height=2.45cm]{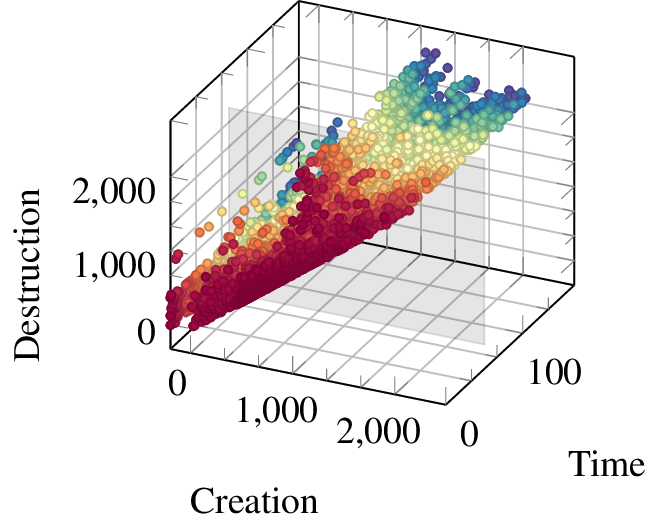}
  }%
  \subcaptionbox{Persistence diagram\label{sfig:Persistence diagram slice}}{%
    \includegraphics[height=2.45cm]{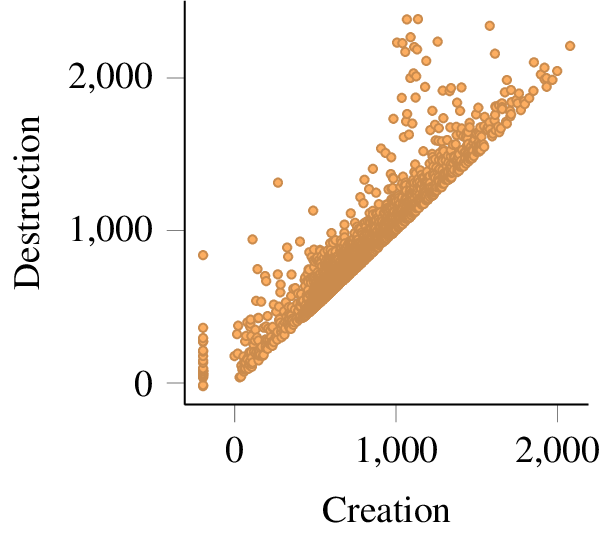}
  }%
  \quad%
  \subcaptionbox{Summary statistics\label{sfig:Summary statistics}}{%
    \begin{tikzpicture}
      \begin{groupplot}[%
        group style = {%
          group size     = 1 by 2,
          vertical sep   = 0pt,
          xticklabels at = edge bottom,
        },
        axis x line*     = bottom,
        axis y line*     = left,
        enlarge x limits = false,
        enlarge y limits = 0.2,
        width            = 0.45\linewidth,
        height           = 2.5cm,
        tick align       = outside,
        scaled ticks     = false,
        ticklabel style  = {font = \tiny},
        label style      = {font = \tiny},
        %
        % Do *not* convert this to math mode
        tick label style = {
          /pgf/number format/fixed,
          /pgf/number format/assume math mode = true
        },
        colormap/Spectral,
      ]
        \nextgroupplot
          \addplot[mesh, scatter src = x, draw = black, thick, no marks] table[col sep=comma, x=time, y=infinity_norm_p1]{./Data/Summary_statistics_042.csv};
          \addplot[scatter, scatter src = x, thick, no marks, mesh] table[col sep=comma, x=time, y=infinity_norm_p1]{./Data/Summary_statistics_042.csv};
          \draw[thick, opacity = 0.1] (50, 0) -- (50, 1500);

          \node[anchor=west] at (rel axis cs:0.10, 0.15) {\tiny $\|\diagram\|_\infty$};
        \nextgroupplot
          \addplot[mesh, scatter src = x, draw = black, thick, no marks] table[col sep=comma, x=time, y=total_persistence_p1]{./Data/Summary_statistics_042.csv};
          \addplot[scatter, scatter src = x, thick, no marks, mesh] table[col sep=comma, x=time, y=total_persistence_p1]{./Data/Summary_statistics_042.csv};
          \draw[thick, opacity = 0.1] (50, 0) -- (rel axis cs:0.2935, 1.0);

          \node[anchor=west] at (rel axis cs:0.10, 0.15) {\tiny $\|\diagram\|_1$};
      \end{groupplot}
    \end{tikzpicture}
  }
  \caption{%
    Example of summary statistics calculations. Starting from a sequence
    of time-varying persistence diagrams~\subref{sfig:Persistence diagram
    3D} of one participant, for
    each diagram slice~\subref{sfig:Persistence diagram slice}, we evaluate a scalar-valued
    statistic~$S\colon\diagram\to\reals$, leading to a time
    series~\subref{sfig:Summary statistics}; the corresponding time
    point is highlighted.
  }
  \label{fig:Summary statistics example}
\end{figure}
%%%%%%%%%%%%%%%%%%%%%%%%%%%%%%%%%%%%%%%%%%%%%%%%%%%%%%%%%%%%%%%%%%%%%%%%

Extracting information from the time-varying persistence diagrams of
each participant is impeded by their complex geometrical
structure, making it necessary to use summary
statistics.
We first focus on a description of \emph{global} properties of
participants, restricting ourselves to persistence diagrams with $d
= 2$~(i.e.\ we are studying voids of the activation function).
To this end, we calculate topological summary statistics of the form
$S\colon\diagram\to\reals$.
We calculate two related summary statistics here, namely
the \emph{infinity norm} $\|\diagram\|_\infty$ of a persistence
diagram~\citep{Cohen-Steiner07} and the \emph{$p$-norm}\footnote{%
  The term \emph{total persistence} is sometimes used interchangeably for this norm.
}
$\|\diagram\|_p$~\citep{Cohen-Steiner10, Chen11}, defined by
\begin{equation}
  \|\diagram\|_\infty := \max_{x,y \in \diagram} \persistence(x, y)^{p} \quad\text{and}\quad\|\diagram\|_p := \sqrt[\leftroot{-1}\uproot{3}p]{\sum_{x, y \in \diagram} \persistence(x, y)^{p}},
  \label{eq:Summary statistics norm}
\end{equation}
with $p \in \reals$. We found $p = 1$ to be sufficient, thus
using unscaled persistence values. Since both norms in \autoref{eq:Summary
statistics norm} yield one scalar value for a persistence
diagram, the summary statistics turn a sequence of time-varying
persistence diagrams into a time series of scalar-valued summary
statistics. \autoref{fig:Summary statistics example} depicts this for
a single participant from our data~(for illustrative purposes, we show
\emph{all} 168 time steps; as specified before, only 161 time steps will
be used for the subsequent analyses).

%%%%%%%%%%%%%%%%%%%%%%%%%%%%%%%%%%%%%%%%%%%%%%%%%%%%%%%%%%%%%%%%%%%%%%%%

% Don't tell anyone we are doing this. To put it more dramatically:
%
%   lasciate ogne speranza, voi ch'intrate
%
% Seriously, though, this is slightly hacky.
\newsavebox{\UL}
\newsavebox{\UR}
\newsavebox{\LL}
\newsavebox{\LR}

\begin{figure}
  \centering
  \sbox{\UL}{\subcaptionbox{\textsc{\scriptsize{}baseline-tt}}{%
      \begin{tikzpicture}
        \begin{axis}[%
          width              = 4.0cm,
          height             = 4.0cm,
          axis lines         = none,
          colormap/Set1-6,
          colormap = {reverse set}{
            indices of colormap={
                \pgfplotscolormaplastindexof{Set1-6},...,0 of Set1-6}
          },
          point meta min  = 0,
          point meta max  = 5,
          ticklabel style = {font = \small},
          label style     = {font = \small},
          mark size       = 1pt,
          % Just to make the visualisations align nicely to each other.
          y dir           =  reverse,
        ]
          \addplot[%
            scatter,
            only marks,
            point meta = explicit]
              table[col sep = comma,
                    x       = x,
                    y       = y,
                    meta    = cohort] {Data/embeddings/baseline_autocorrelation_brainmask.csv};
        \end{axis}
      \end{tikzpicture}
    }%
  }%
  \sbox{\UR}{\subcaptionbox{\textsc{\scriptsize{}baseline-pp}}{%
      \begin{tikzpicture}
        \begin{axis}[%
          width              = 4.0cm,
          height             = 4.0cm,
          axis lines         = none,
          colormap/Set1-6,
          colormap = {reverse set}{
            indices of colormap={
                \pgfplotscolormaplastindexof{Set1-6},...,0 of Set1-6}
          },
          point meta min  = 0,
          point meta max  = 5,
          ticklabel style = {font = \small},
          label style     = {font = \small},
          mark size       = 1pt,
          % Just to make the visualisations align nicely to each other.
          y dir           =  reverse,
        ]
          \addplot[%
            scatter,
            only marks,
            point meta = explicit]
              table[col sep = comma,
                    x       = x,
                    y       = y,
                    meta    = cohort] {./Data/embeddings/baseline_autocorrelation_parcellated_brainmask.csv};
        \end{axis}
      \end{tikzpicture}
    }%
  }
  \sbox{\LL}{\subcaptionbox{$\|\diagram\|_1$}{%
      \begin{tikzpicture}
        \begin{axis}[%
          width              = 4.0cm,
          height             = 4.0cm,
          axis lines         = none,
          colormap/Set1-6,
          colormap = {reverse set}{
            indices of colormap={
                \pgfplotscolormaplastindexof{Set1-6},...,0 of Set1-6}
          },
          point meta min  = 0,
          point meta max  = 5,
          ticklabel style = {font = \small},
          label style     = {font = \small},
          mark size       = 1pt,
          % Just to make the visualisations align nicely to each other.
          y dir           =  reverse,
        ]
          \addplot[%
            scatter,
            only marks,
            point meta = explicit]
              table[col sep = comma,
                    x       = x,
                    y       = y,
                    meta    = cohort] {Data/embeddings/brainmask_total_persistence_p1.csv};
        \end{axis}
      \end{tikzpicture}
    }%
  }%
  \sbox{\LR}{%
    \subcaptionbox{$\|\diagram\|_\infty$}{%
      \begin{tikzpicture}
        \begin{axis}[%
          width              = 4.0cm,
          height             = 4.0cm,
          axis lines         = none,
          colormap/Set1-6,
          colormap = {reverse set}{
            indices of colormap={
                \pgfplotscolormaplastindexof{Set1-6},...,0 of Set1-6}
          },
          point meta min  = 0,
          point meta max  = 5,
          ticklabel style = {font = \small},
          label style     = {font = \small},
          mark size       = 1pt,
        ]
          \addplot[%
            scatter,
            only marks,
            point meta = explicit]
              table[col sep = comma,
                    x       = x,
                    y       = y,
                    meta    = cohort] {Data/embeddings/brainmask_infinity_norm_p1.csv};
        \end{axis}
      \end{tikzpicture}
    }%
  }
  \begin{minipage}[b]{0.5\linewidth}
    \usebox{\UL}
    \usebox{\UR}\\
    \usebox{\LL}
    \usebox{\LR}%
  \end{minipage}%
  \subcaptionbox{Age prediction task\label{sfig:Age prediction}}{%
    \small
    \robustify\bfseries
    \begin{tabular}{lSSS}
      \toprule
      Method & BM & OM & XM\\
      \midrule
      \textsc{baseline-tt}              &           0.09 &          0.02  &           0.24\\
      \textsc{baseline-pp}              &           0.41 &          0.40  &           0.40\\
      \textsc{tt-corr-tda}              &           0.17 &           0.11 &           0.23\\
      \textsc{pp-corr-tda}              &           0.25 &           0.27 &           0.23\\
      \midrule
      \textsc{srm}                      &           0.44 &         {---}  &          {---}\\
      \midrule
      $\|\diagram\|_1$                  &           0.46 &           0.67 &           0.48\\
      $\|\diagram\|_1$ parcellated      &           0.32 &           0.50 &           0.34\\
      $\|\diagram\|_\infty$             &           0.61 & \bfseries 0.77 & \bfseries 0.73\\
      $\|\diagram\|_\infty$ parcellated & \bfseries 0.67 &           0.50 &           0.33\\
      \bottomrule
    \end{tabular}
  }
  \caption{%
    An embedding of the \emph{distances} for different baselines and topological
    summaries, based on the whole-brain mask~(BM); colour-coding refers to the
    age group of participants. The table depicts the results of the age
    prediction task, stratified by different brain masks; performance is
    measured as a correlation coefficient~(bold indicates the best
    results).
  }
  \label{fig:Summary statistics}
\end{figure}
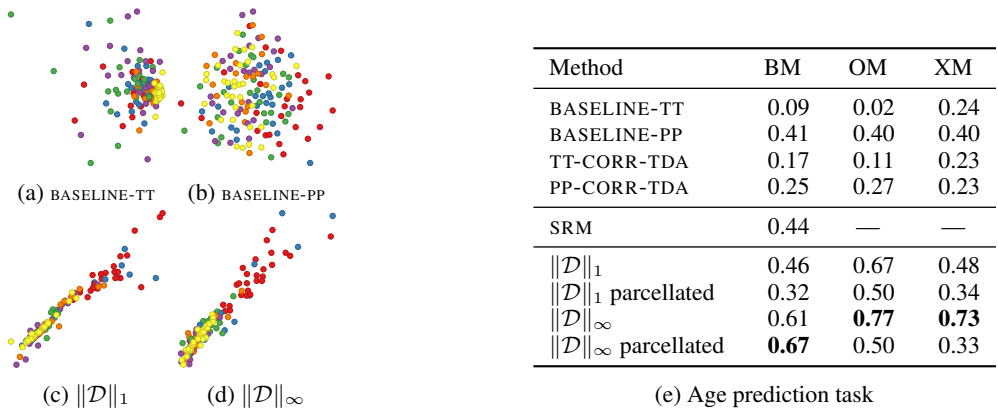
%%%%%%%%%%%%%%%%%%%%%%%%%%%%%%%%%%%%%%%%%%%%%%%%%%%%%%%%%%%%%%%%%%%%%%%%

%%%%%%%%%%%%%%%%%%%%%%%%%%%%%%%%%%%%%%%%%%%%%%%%%%%%%%%%%%%%%%%%%%%%%%%%
\paragraph{Qualitative evaluation.}
%%%%%%%%%%%%%%%%%%%%%%%%%%%%%%%%%%%%%%%%%%%%%%%%%%%%%%%%%%%%%%%%%%%%%%%%
%
\autoref{fig:Summary statistics} shows an embedding obtained from
our topological summary statistics~(using multidimensional scaling based on
the Euclidean distance between per-participant curves) compared to
baseline embeddings, which we obtain from the two correlation
matrices described in \autoref{sec:Related work}. We refer to them as 
\textsc{baseline-tt}~(time-based) and
\textsc{baseline-pp}~(voxel-based; parcellated for computational ease),
respectively~(see \autoref{sec:Baselines} for additional details). 
Both topology-based embeddings are showing a split between
participants. By colour-coding the age group of each participant, we see
that topology-based embeddings separate adults~(red)
from children~(other colours). The baselines, by contrast, do not
exhibit such a clear-cut distinction.

%%%%%%%%%%%%%%%%%%%%%%%%%%%%%%%%%%%%%%%%%%%%%%%%%%%%%%%%%%%%%%%%%%%%%%%%
\paragraph{Quantitative evaluation.}
%%%%%%%%%%%%%%%%%%%%%%%%%%%%%%%%%%%%%%%%%%%%%%%%%%%%%%%%%%%%%%%%%%%%%%%%
%
To \emph{quantify} the benefits of our proposed topological feature extraction
pipeline, we set up a task in which we predict the age of the
non-adult participants.
Using a ridge regression and leave-one-out cross-validation~(see
\autoref{sec:Baselines} for detailed descriptions of all comparison
partners and \autoref{sec:Age prediction} for additional experimental
details), we train models on either the curves of summary
statistics~(not the embeddings) the baseline matrices, and additional
topological baselines, reporting the correlation coefficient in the
table in \autoref{fig:Summary statistics}.
Higher values indicate that the model is
better suited to predict the age. The SRM result comes from previous work on the same data
set~\citep{yates_emergence_2020}; we note that our task is slightly
different\footnote{%
  \citet{yates_emergence_2020} learn a shared set of features in
  adult participants to predict the age of non-adults.
}. Overall, we observe strong correlations, indicating that topological features are highly
useful for age prediction and carry salient information.
Performance based on the occipital-temporal
mask~(OM) and on the XOR mask~(XM) is higher than for the whole-brain
mask~(BM); we hypothesise that this is partially due to the higher noise
level of BM, whereas OM and XM focus only on a subset of the
brain~(which decreases the noise level).
We also note that $\|\diagram\|_{\infty}$, which only
considers the most persistence topological feature of a persistence
diagram, performs best in the prediction task, possibly because it is
more robust to small-scale noise.
Interestingly, parcellated data~(i.e.\ highly coarse
representations) applied to our cubical complex filtration outperforms
the whole-brain mask. This is the \emph{only} one of the parcellated
volumes to do so. We speculate that the coarsening helps to remove some
noise here, whereas the other masks, containing fewer voxels, are less
noisy by construction and contain more fine-grained information that is
suppressed by the coarsening.

%%%%%%%%%%%%%%%%%%%%%%%%%%%%%%%%%%%%%%%%%%%%%%%%%%%%%%%%%%%%%%%%%%%%%%%%
\subsection{Dynamic analysis based on brain state trajectories}\label{sec:Dynamic analysis}
%%%%%%%%%%%%%%%%%%%%%%%%%%%%%%%%%%%%%%%%%%%%%%%%%%%%%%%%%%%%%%%%%%%%%%%%

So far, we dealt only with overall summary statistics. Our framework also
enables analysing the brain state of participants over time.
We sidestep the aforementioned issue of persistence diagram metric
computations by calculating \emph{persistence images}~\citep{Adams17}
from the persistence diagrams. A persistence image is a function
$\Psi\colon\reals^2\to\reals$ that turns a diagram~$\diagram$ into
a surface via $\Psi(z) := \sum_{x, y \in \diagram}
\weight(x, y) \Phi(x, y, z)$, where $\weight(\cdot)$ is a fixed piecewise
linear weight function and $\Phi(\cdot)$ denotes a probability distribution, which is
typically chosen to be a normalised symmetric Gaussian. By
discretising~$\Psi$~(using an $r \times r$ grid), a persistence
diagram is transformed into an image\footnote{%
  Intuitively, this can also be seen as a form of kernel density
  estimation on persistence diagrams.
}; this is depicted in
\autoref{sfig:Persistence images}. The main advantage of~$\Psi$ lies in
embedding persistence diagrams into a space that is amenable to standard
machine learning tools; moreover, $\Psi$ affords defining and calculating
\emph{unique} means, as opposed to persistence diagrams~\citep{Munch13,
Munch15, Turner14}.
Subsequently, we use $r = 20$ and a Gaussian kernel with $\sigma = 1.0$;
$\Psi$ is known to be impervious to such choices~\citep{Adams17}.

%%%%%%%%%%%%%%%%%%%%%%%%%%%%%%%%%%%%%%%%%%%%%%%%%%%%%%%%%%%%%%%%%%%%%%%%
\subsubsection{Cohort brain state trajectories}
%%%%%%%%%%%%%%%%%%%%%%%%%%%%%%%%%%%%%%%%%%%%%%%%%%%%%%%%%%%%%%%%%%%%%%%%

%%%%%%%%%%%%%%%%%%%%%%%%%%%%%%%%%%%%%%%%%%%%%%%%%%%%%%%%%%%%%%%%%%%%%%%%
\begin{figure}[tbp]
  \centering
  \subcaptionbox{Whole-brain mask~(entropy: 0.61, 0.97, 1.73, 1.15, 1.30, 1.67)\label{sfig:Trajectories BM}}{%
    \includegraphics[width=0.875\linewidth]{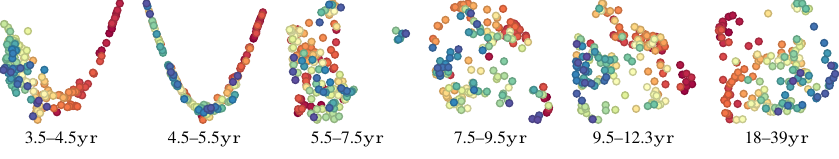}
  }
  \vspace{0.5\baselineskip}
  \subcaptionbox{Occipital-temporal mask~(entropy: 0.87, 0.65, 1.03, 0.94, 0.82, 1.46)\label{sfig:Trajectories OM}}{%
    \includegraphics[width=0.875\linewidth]{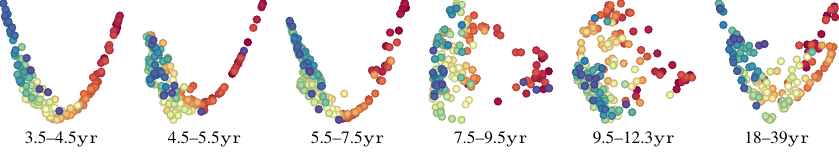}
  }
  \vspace{0.5\baselineskip}
  \subcaptionbox{XOR mask~(entropy: 0.86, 0.85, 1.13, 0.60, 0.88, 0.87)\label{sfig:Trajectories XM}}{%
    \includegraphics[width=0.875\linewidth]{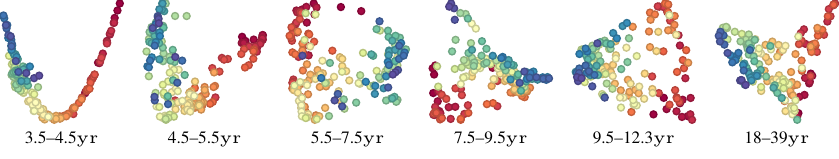}
  }
  \caption{%
    Cohort brain state trajectories for different brain masks, embedded
    using PHATE~\citep{Moon19}. Annotations provide the age range of
    subjects in one cohort. We also report the von Neumann entropy of
    the respective diffusion operator~\citep{Anand11}.
  }
  \label{fig:Brain state trajectories}
\end{figure}
%%%%%%%%%%%%%%%%%%%%%%%%%%%%%%%%%%%%%%%%%%%%%%%%%%%%%%%%%%%%%%%%%%%%%%%%

By evaluating~$\Psi(\diagram_2^{(i, j)})$ for each time step~$t_j$, we
turn the sequence of persistence diagrams of the $i$th participant
into a matrix~$\featurematrix^{(i)} \in \reals^{m \times r^2}$, where
the $j$th row corresponds to the `unravelled' persistence image of
time step~$t_j$. We now calculate the sample mean $\overline{\featurematrix}_k$
of each participant cohort, resulting in six matrices whose rows
represent the average topological activity of participants in the
respective cohort. Taking the Euclidean distance between persistence
images as a proxy for their actual topological
dissimilarity~\citep[Theorem~3]{Adams17},
we calculate pairwise distances between rows of each
$\overline{\featurematrix}_k$ and embed them using
PHATE~\citep{Moon19}, a powerful embedding algorithm for time-varying
data. This turns $\overline{\featurematrix}_k$ into a~2D
\emph{brain state trajectory}~(where the state is measured using
topological features). \autoref{fig:Brain state trajectories} depicts
the resulting trajectories for different masks.
All brain state trajectories exhibit visually distinct behaviour in older and
younger subjects. The youngest subjects are characterised by a simple
`linear' trajectory in the whole-brain mask, indicating that
their processing of the movie is more sensory-driven. This pattern is
visible in \autoref{sfig:Trajectories OM} for young
children in general: until \SI{7.5}{\year}, sensory processing, analysed
using the occipital-temporal mask, is comparatively simple. In older
subjects, we observe more complex trajectories with higher entropy
generally. Developmental differences are best indicated in
\autoref{sfig:Trajectories XM}, where we see that the overall trajectory
shape becomes `adult-like' \emph{earlier}~(and thus more complex).
Since this mask is composed of more cognitive brain regions~(rather
than sensory ones), we hypothesise that this could indicate that older
participants, including older children, are capable of connecting
different aspects of the movie to their memories, for example, whereas
the simpler trajectories of the two youngest cohorts in \emph{all} brain
masks may indicate that these participants are not comprehending the
movie on a non-superficial level.

%%%%%%%%%%%%%%%%%%%%%%%%%%%%%%%%%%%%%%%%%%%%%%%%%%%%%%%%%%%%%%%%%%%%%%%%
\subsubsection{Variability analysis}\label{sec:Variability analysis}
\begin{figure}[tbp]
  \centering
  \subcaptionbox{Whole-brain mask}{%
    \includegraphics[width=0.25\linewidth]{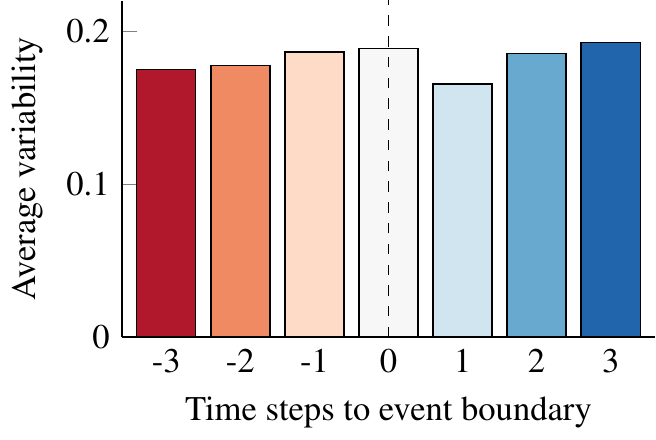}%
  }%
  \vspace{0.5\baselineskip}%
  \subcaptionbox{Occipital-temporal mask}{%
    \includegraphics[width=0.25\linewidth]{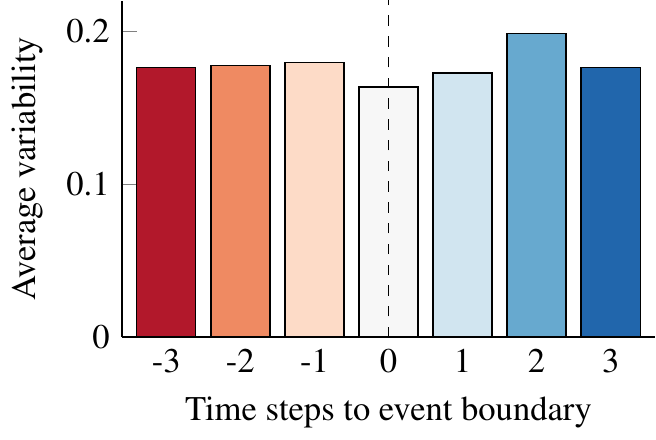}%
  }%
  \vspace{0.5\baselineskip}%
  \subcaptionbox{XOR mask}{%
    \includegraphics[width=0.25\linewidth]{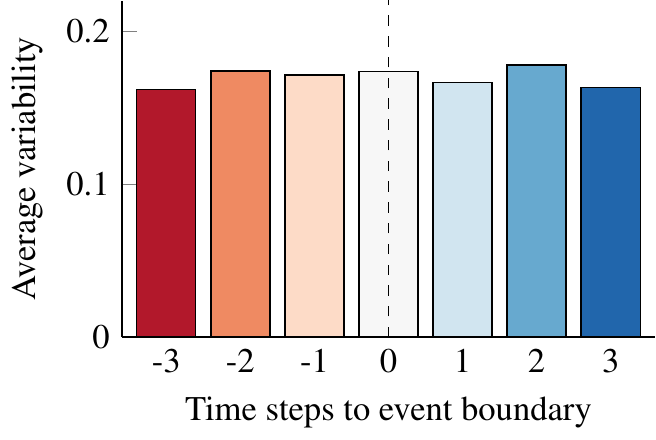}%
  }%
  \caption{%
    Histograms showing the mean across-cohort variability as a function
    of the distance to an event. The $x$-axis shows the time steps
    prior to~(negative) or after~(positive) an event boundary, while their
    $y$-axis depicts across-cohort variability. Please refer to
    \autoref{sec:Variability analysis} for more details.
  }
  \label{fig:Variability histograms}
\end{figure}

%%%%%%%%%%%%%%%%%%%%%%%%%%%%%%%%%%%%%%%%%%%%%%%%%%%%%%%%%%%%%%%%%%%%%%%%
%
To quantify the variability across cohorts, we
calculate the per-column maximum of each~$\overline{\featurematrix}_k$,
referring to the respective set of values as
$\|\overline{\featurematrix}_k\|_{\infty}\in\reals^m$; the calculated
values are the equivalent of the infinity norm evaluated~(per time step)
on a mean persistence image of the cohort.
We finally calculate
$s\left(\|\overline{\featurematrix}_1\|_{\infty}, \dots,
  |\overline{\featurematrix}_6\|_{\infty}\right)$, i.e.\ the sample
  standard deviation per time point, thus obtaining a \emph{variability
  curve} of~$m$ time steps~(see \autoref{fig:Across-cohort variability
curves}).
To use this variability curve, we ran an online study to discover which
salient events are detected by participants in the movie. Using 22~test
subjects~(with no overlap to the ones used in the fMRI data acquisition
process), we followed \citet{Ben-Yakov18} and determined consensus
boundaries of events in the movie. We declare an event boundary to be \emph{salient} if 
at least~7 participants agree, resulting in~20 events.
Given this information, we collect the average variability over all
events for a window of $w = 3$ time steps before and after an event,
leading to averaged variabilities $\{s_1, \dots, s_7\}$, where $s_4$
corresponds to the average variability at the event boundary itself~(see
\autoref{fig:Variability histograms}).
It is our hypothesis that post-event and pre-event variability are
different---in other words, our topological features capture cognitive
differences across cohorts and events. To quantify this, we calculate
$s_{\text{pre}} := \max_{i \leq 3} s_i - \min_{i \leq 3} s_i$
and
$s_{\text{post}} := \max_{i \geq 5} s_i - \min_{i \geq 5} s_i$.
We set $\theta := s_{\text{pre}} - s_{\text{post}}$ as our test
statistic and perform a bootstrap procedure by sampling~20 time
points at random and repeating the same calculation, thereby obtaining
an empirical null distribution.
This results in bootstrap samples $\widehat{\theta}_1, \dots,
\widehat{\theta}_{1000}$ serving as a null
distribution~$\widehat{\theta}$, from which we obtain the achieved
significance level~(ASL) as $\probability(\widehat{\theta} \geq
\theta)$.

The ASL values are $0.084$~(whole-brain mask, BM),
$0.045$~(occipital-temporal mask, OM), and $0.396$~(XOR mask, XM),
respectively, indicating that the effect of capturing events is \emph{strongest}
in OM and significant at the $\alpha = 0.05$ level.
This aligns well with the gradual differences between cohorts expressed
in \autoref{sfig:Trajectories OM}. Event differences are less pronounced in
BM~(which, as \autoref{sfig:Trajectories BM} shows,
is capturing more complex cohort patterns). Finally, event differences
are \emph{absent} in XM, showing that across-cohort
variability is not consistent with event boundaries here, hinting at the
fact that this mask might better be used to assess within-cohort
variability rather than across-cohort variability.
Please refer to \autoref{sec:Across-cohort variability analysis} for
additional visualisations.

%%%%%%%%%%%%%%%%%%%%%%%%%%%%%%%%%%%%%%%%%%%%%%%%%%%%%%%%%%%%%%%%%%%%%%%%
%%%%%%%%%%%%%%%%%%%%%%%%%%%%%%%%%%%%%%%%%%%%%%%%%%%%%%%%%%%%%%%%%%%%%%%%

%%%%%%%%%%%%%%%%%%%%%%%%%%%%%%%%%%%%%%%%%%%%%%%%%%%%%%%%%%%%%%%%%%%%%%%%
\section{Conclusion}
%%%%%%%%%%%%%%%%%%%%%%%%%%%%%%%%%%%%%%%%%%%%%%%%%%%%%%%%%%%%%%%%%%%%%%%%

This paper demonstrates the potential of an unsupervised, non-parametric
topology-based feature extraction framework for fMRI data, permitting
both static and dynamic analyses. We showed that topological summary
statistics are useful in an age prediction task. Using vectorised
topological features descriptors, we also developed cohort brain state
trajectories that show the time-varying behaviour of a cohort of
participants~(binned by age). Next, to highlight qualitative
age-related differences in the overall cognition of participants, we
were also able to uncover quantitative differences in event
processing.
In the future, we want to further analyse the \emph{geometry} of brain
state trajectories and link states back to events; a preliminary
analysis~(see \autoref{sec:Curvature analysis}) finds significant
differences between the mean curvature~\citep{doCarmo76} of adult and
non-adult participants, thus showcasing the explanatory potential of
topological features. We also plan on investigating geometrical
aspects of topological features~\citep{Erickson05, Zomorodian08} 
as well as their large-scale validation based on synthetic data
generators~\citep{Ellis20}.

%%%%%%%%%%%%%%%%%%%%%%%%%%%%%%%%%%%%%%%%%%%%%%%%%%%%%%%%%%%%%%%%%%%%%%%%
% Impact statement
%%%%%%%%%%%%%%%%%%%%%%%%%%%%%%%%%%%%%%%%%%%%%%%%%%%%%%%%%%%%%%%%%%%%%%%%

\section*{Broader impact}

The primary contribution of this work---a novel, parameter-free way of extracting informative
features from fMRI data---is of a computational nature.
In general, we fully acknowledge that any researcher dealing with fMRI
data analysis~(not necessarily restricted to machine learning methods)
has a big responsibility.
Since our work is purely computational,
we do not believe that it will have adverse ethical consequences,
provided the experimental design is unbiased. For the same reason,
our work is not specifically favouring or disfavouring any groups.

Beyond the immediate applications for fMRI data analysis, our work also
has a broader applicability for the analysis of time-varying or
structured neuroscience data in general. This includes other non-invasive
techniques such as EEG or MEG, but also neuronal spike data from cell
populations. Our work is appealing for such data because it does not
\emph{require} auxiliary representations such as graphs.
We are thus convinced that the introduction of our directly-computable
topological features will overall have beneficial outcomes.

As long-term goal, for example, our work could serve as a foundation to
investigate neurological pathologies~(such as depressive disorders) from
a new, topological perspective. In general, our dynamic analyses also
allow us to capture not just stable traits in different populations, but
also the different mental states participants progress through while
undergoing fMRI. As a generic feature descriptor of brain states, we
would welcome a future in which topological features aid in
understanding such traits or states.

\begin{ack}
We thank the anonymous reviewers for their valuable comments, which
helped us improve the paper.
The first author is also indebted to Michael Moor, Leslie O'Bray, and
Caroline Weis for their constructive feedback in preparing the manuscript.
This work was partially funded and supported by the Swiss National
Science Foundation~[Spark grant~190466, \emph{B.R.}],
the Alfried Krupp Prize for Young University Teachers of the Alfried
Krupp von Bohlen und Halbach-Stiftung~[\emph{K.B.}],
IVADO Professor startup \& operational funds~[\emph{G.W.}], 
an NSF Graduate Research Fellowship [\emph{T.Y.}], 
NSF grant CCF 1839308 \& Canadian Institute for Advanced Research [\emph{N.T.B.}], 
Chan-Zuckerberg Initiative grants 182702 \& CZF2019-002440~[\emph{S.K.}], 
and NIH grants R01GM135929 \& R01GM130847~[\emph{G.W., S.K.}].
The content provided here is solely the responsibility of the authors and 
does not necessarily represent the official views of the funding agencies. 
The funders had no role in study design, data collection \& analysis, 
decision to publish, or preparation of the manuscript.
\end{ack}

%%%%%%%%%%%%%%%%%%%%%%%%%%%%%%%%%%%%%%%%%%%%%%%%%%%%%%%%%%%%%%%%%%%%%%%%
% References
%%%%%%%%%%%%%%%%%%%%%%%%%%%%%%%%%%%%%%%%%%%%%%%%%%%%%%%%%%%%%%%%%%%%%%%%

{
  % Makes the preprint look nicer; we are allowed to do this according to
  % the style guide.
  \small

  \bibliographystyle{abbrvnat}
  \bibliography{main}
}

%%%%%%%%%%%%%%%%%%%%%%%%%%%%%%%%%%%%%%%%%%%%%%%%%%%%%%%%%%%%%%%%%%%%%%%%
% Appendix
%%%%%%%%%%%%%%%%%%%%%%%%%%%%%%%%%%%%%%%%%%%%%%%%%%%%%%%%%%%%%%%%%%%%%%%%

\clearpage
\appendix

% This ensures that we get a prefix that depends on the section of the
% current document, i.e. the appendix in this case.
\counterwithin{figure}{section}
\counterwithin{table} {section}

%%%%%%%%%%%%%%%%%%%%%%%%%%%%%%%%%%%%%%%%%%%%%%%%%%%%%%%%%%%%%%%%%%%%%%%%
\section{Appendix}
%%%%%%%%%%%%%%%%%%%%%%%%%%%%%%%%%%%%%%%%%%%%%%%%%%%%%%%%%%%%%%%%%%%%%%%%

The following sections provide additional details about the experiments
as well as a brief glimpse into other analyses that we are actively
pursuing for future work.

%%%%%%%%%%%%%%%%%%%%%%%%%%%%%%%%%%%%%%%%%%%%%%%%%%%%%%%%%%%%%%%%%%%%%%%%
\subsection{An intuitive introduction to persistent homology}\label{sec:Persistent homology}
%%%%%%%%%%%%%%%%%%%%%%%%%%%%%%%%%%%%%%%%%%%%%%%%%%%%%%%%%%%%%%%%%%%%%%%%

Persistent homology was developed as a `shape descriptor' for real-world
data sets, where the idealised notions of algebraic topology do not
necessarily apply any more. This is illustrated by the subsequent
figure, which deals with a point cloud that has a roughly circular
shape. Notice that this shape is immediately recognisable to humans,
but from the perspective of algebraic topology, it is merely
a collection of points with a trivial shape.

%%%%%%%%%%%%%%%%%%%%%%%%%%%%%%%%%%%%%%%%%%%%%%%%%%%%%%%%%%%%%%%%%%%%%%%%
\begin{figure}[h]
{
  \centering
  \pgfplotsset{
    every axis/.append style = {
      axis lines         = none,
      mark               = *,
      mark size          = 0.5pt,
      unit vector ratio* = 1 1 1,
      width              = 4.0cm,
      height             = 4.0cm,
    },
  }
  \begin{subfigure}{0.25\textwidth}%
    \centering
    \begin{tikzpicture}
      \begin{axis}[
      ]
        \addplot[black, only marks] table {Data/Circle.txt};
      \end{axis}
    \end{tikzpicture}
  \end{subfigure}%
  \begin{subfigure}{0.25\textwidth}%
    \centering
    \begin{tikzpicture}
      \begin{axis}[
      ]
        \addplot[black, mark=*] table {Data/Circle_0_1.txt};
      \end{axis}
    \end{tikzpicture}
  \end{subfigure}%
  \begin{subfigure}{0.25\textwidth}%
    \centering
    \begin{tikzpicture}
      \begin{axis}[
      ]
        \addplot[black, mark=*] table {Data/Circle_0_2.txt};
      \end{axis}
    \end{tikzpicture}
  \end{subfigure}%
  \begin{subfigure}{0.25\textwidth}
    \centering
    \begin{tikzpicture}
      \begin{axis}[
      ]
        \addplot[black, mark=*] table {Data/Circle_0_3.txt};
      \end{axis}
    \end{tikzpicture}
  \end{subfigure}
}
\end{figure}
%%%%%%%%%%%%%%%%%%%%%%%%%%%%%%%%%%%%%%%%%%%%%%%%%%%%%%%%%%%%%%%%%%%%%%%%

We observe that we \emph{can} analyse this point cloud by picking an
appropriate scale parameter. More precisely, if we start connecting
points that are within a certain distance~$\epsilon$ to each other, we
obtain a nested sequence of simplicial complexes~(in our context, this
term is synonymous with a graph) as we increase~$\epsilon$. This is
a special type of filtration---a filtration based on pairwise distances,
and the resulting simplicial complexes are depicted above. It is now
possible to calculate Betti numbers for each of these complexes. Since
we are only dealing with~2D points, there are only two relevant Betti
numbers, namely~$\betti{0}$ and~$\betti{1}$, corresponding to the number
of connected components and the number of cycles, respectively. Suppose
now that we \emph{track} these numbers for each one of the steps in the
filtration; moreover, suppose we have a way of making the individual
steps in the filtration as small as possible such that we never miss
any changes in~$\betti{0}$ and~$\betti{1}$. For every topological
feature---every component and every cycle---we can thus measure
precisely when a feature was created and when it was destroyed.

\paragraph{Persistence diagram.}
This information is collected in the persistence diagram, which
summarises all topological activity. 
In this example, the persistence diagram of the \mbox{1-dimensional}
topological features contains a few points, each one of them
corresponding to one specific cycle in the data.
%
%%%%%%%%%%%%%%%%%%%%%%%%%%%%%%%%%%%%%%%%%%%%%%%%%%%%%%%%%%%%%%%%%%%%%%%%
\begin{wrapfigure}[15]{r}[0pt]{4cm}
  \begin{tikzpicture}
    \begin{axis}[%
      axis x line*       = bottom,
      axis y line*       = left,
      mark size          = 0.5pt,
      tick align         = outside,
      unit vector ratio* = 1 1 1,
      xmin               = 0.0,
      ymin               = 0.0,
      xmax               = 1.0,
      ymax               = 1.0,
      width              = 5cm,
      ticklabel style  = {font = \tiny},
      label style      = {font = \tiny},
      % Do *not* convert this to math mode
      tick label style = {
        /pgf/number format/assume math mode = true
      },
    ]
      \addplot[only marks] table {Data/Circle_0_3_PD.txt};
      \addplot[domain={0:1}] {x};
    \end{axis}
  \end{tikzpicture}
  \caption*{
    A persistence diagram of the \mbox{1-dimensional} topological
    features~(cycles).
  }
\end{wrapfigure}
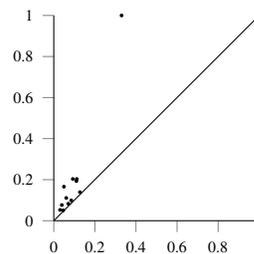
%%%%%%%%%%%%%%%%%%%%%%%%%%%%%%%%%%%%%%%%%%%%%%%%%%%%%%%%%%%%%%%%%%%%%%%%
%
The axes correspond to the scale
parameter~(their actual values can be safely ignored for this illustrative example). The
\mbox{$x$-axis} shows the threshold at which a cycle was \emph{created},
i.e.\ at which there is a `hole' in the corresponding simplicial
complex, while the \mbox{$y$-axis} depicts the threshold at which this
hole is \emph{destroyed}, i.e.\ closed. We do not specifically indicate
this here, but cycles are destroyed whenever all points that are
involved in their creation are connected to each other. Put differently,
this means that we ignore cycles created by individual triangles of
points, for example, as they are qualitatively different from cycles
created by arranging points in such a circular shape~(there are more
technical reasons for this restriction). In any case, the persistence
diagram demonstrates that virtually all cycles---depicted as
points---occur at small scales, except for \emph{one}.
This coincides with our intuition: we do not perceive such a point
cloud to have a lot of large-scale cycles. The persistence diagram thus
serves as an intuitive feature descriptor: points that occur at large
scales are far removed from the diagonal~(and have a high persistence),
whereas the small-scale features cluster around the diagonal.

The interesting fact is that knowing the persistence diagram also
makes it possible for us to `guess' the number of relevant scales of
a point cloud! In this example, we would possibly state that there is
only \emph{one} useful scale at which to analyse the data, namely the
scale for which the cycle structure becomes topologically apparent. In
general, this will differ based on the data set. Persistent homology
does not force us to prefer a scale, making it suitable for the analysis
of real-world data sets. The ingenious realisation of
\citet{Edelsbrunner02} was that there is no reason to `guess' scales or
compute Betti numbers per step, as we described it above. Instead, it is
possible to obtain information about \emph{all} potential scales by
a single pass through the data, making this a highly-efficient
algorithm~(at least as long as the dimension of the input data is
bounded; calculating topological features for dimensions $d \gg 3$
efficiently is still a topic of ongoing research).

%%%%%%%%%%%%%%%%%%%%%%%%%%%%%%%%%%%%%%%%%%%%%%%%%%%%%%%%%%%%%%%%%%%%%%%%
\paragraph{Persistence images.}
%%%%%%%%%%%%%%%%%%%%%%%%%%%%%%%%%%%%%%%%%%%%%%%%%%%%%%%%%%%%%%%%%%%%%%%%
%
Since the metric structure of persistence diagrams is known to be
complex~\citep{Munch15}, various kernel-based and `vectorisation'
methods exist. In the main text, we focus on \emph{persistence
images}~\citep{Adams17}, a technique that essentially estimates the
density of a persistence diagram and uses a grid to obtain a fixed-size
representation. Such representations may then be used for downstream
processing tasks. As a worked example, consider the following
persistence diagram. After rotating it so that the diagonal becomes the
new \mbox{$x$-axis}, we can perform density estimates with different
resolutions. The density estimator, which is by default a Gaussian
kernel, can be adjusted as well, but \citet{Adams17} mention that this
does \emph{not} have a large influence on the results~(whereas the
resolution should be sufficiently large to capture differences). In the
main paper, we use a smoothing value of~$\sigma = 1$ and a resolution
of~$r = 20$, resulting in \mbox{$400$-dimensional} vectors. We also
calculated different resolutions and smoothing values, but the results
are virtually identical, unless the resolution is decreased too
much: recall that a single persistence diagram of one participant has
around \num{10000} features; reducing them to a, say, $5 \times 5$ image
results in a large loss of information.
%
%%%%%%%%%%%%%%%%%%%%%%%%%%%%%%%%%%%%%%%%%%%%%%%%%%%%%%%%%%%%%%%%%%%%%%%%
\begin{figure}[h]
{
  \centering
  \pgfplotsset{
    every axis/.append style = {
      axis lines         = none,
      mark               = *,
      mark size          = 0.5pt,
      unit vector ratio* = 1 1 1,
      width              = 4.0cm,
      height             = 4.0cm,
    },
  }
  \begin{subfigure}{0.20\textwidth}%
    \centering
    \begin{tikzpicture}
      \begin{axis}[
      ]
        \addplot[black, only marks] table {Data/Persistence_diagram_example_supplement.txt};
      \end{axis}
    \end{tikzpicture}
  \end{subfigure}%
  \begin{subfigure}{0.20\textwidth}%
    \centering
    \begin{tikzpicture}
      \begin{axis}[
      ]
        \addplot[black, only marks] table[y expr = \thisrowno{1} - \thisrowno{0}] {Data/Persistence_diagram_example_supplement.txt};
      \end{axis}
    \end{tikzpicture}
  \end{subfigure}%
  \begin{subfigure}{0.20\textwidth}%
    \centering
    \begin{tikzpicture}
      \begin{axis}[
        enlargelimits = false,
        point meta    = explicit,
        colormap/viridis,
      ]
        \addplot[matrix plot] file {Data/PI_05.txt};
      \end{axis}
    \end{tikzpicture}
  \end{subfigure}%
  \begin{subfigure}{0.20\textwidth}%
    \centering
    \begin{tikzpicture}
      \begin{axis}[
        enlargelimits = false,
        point meta    = explicit,
        colormap/viridis,
      ]
        \addplot[matrix plot] file {Data/PI_10.txt};
      \end{axis}
    \end{tikzpicture}
  \end{subfigure}%
  \begin{subfigure}{0.20\textwidth}
    \centering
    \begin{tikzpicture}
      \begin{axis}[
        enlargelimits = false,
        point meta    = explicit,
        colormap/viridis,
      ]
        \addplot[matrix plot] file {Data/PI_20.txt};
      \end{axis}
    \end{tikzpicture}
  \end{subfigure}
}
\end{figure}
%%%%%%%%%%%%%%%%%%%%%%%%%%%%%%%%%%%%%%%%%%%%%%%%%%%%%%%%%%%%%%%%%%%%%%%%

%%%%%%%%%%%%%%%%%%%%%%%%%%%%%%%%%%%%%%%%%%%%%%%%%%%%%%%%%%%%%%%%%%%%%%%%
\subsection{Properties of cubical complexes}
%%%%%%%%%%%%%%%%%%%%%%%%%%%%%%%%%%%%%%%%%%%%%%%%%%%%%%%%%%%%%%%%%%%%%%%%

%%%%%%%%%%%%%%%%%%%%%%%%%%%%%%%%%%%%%%%%%%%%%%%%%%%%%%%%%%%%%%%%%%%%%%%%
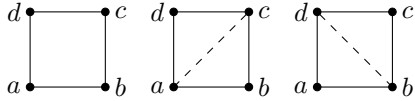
\begin{figure}[t]
  \centering
  \begin{tikzpicture}
    \coordinate[label={[anchor=east]$a$}] (A) at (0, 0);
    \coordinate[label={[anchor=west]$b$}] (B) at (1, 0);
    \coordinate[label={[anchor=west]$c$}] (C) at (1, 1);
    \coordinate[label={[anchor=east]$d$}] (D) at (0, 1);

    \foreach \n in {A, B, C, D}
      \filldraw (\n) circle (0.05cm);

    \draw (A) -- (B) -- (C) -- (D) -- (A);
  \end{tikzpicture}
  \begin{tikzpicture}
    \coordinate[label={[anchor=east]$a$}] (A) at (0, 0);
    \coordinate[label={[anchor=west]$b$}] (B) at (1, 0);
    \coordinate[label={[anchor=west]$c$}] (C) at (1, 1);
    \coordinate[label={[anchor=east]$d$}] (D) at (0, 1);

    \draw[dashed] (A) -- (C);

    \foreach \n in {A, B, C, D}
      \filldraw (\n) circle (0.05cm);

    \draw (A) -- (B) -- (C) -- (D) -- (A);

  \end{tikzpicture}
  \begin{tikzpicture}
    \coordinate[label={[anchor=east]$a$}] (A) at (0, 0);
    \coordinate[label={[anchor=west]$b$}] (B) at (1, 0);
    \coordinate[label={[anchor=west]$c$}] (C) at (1, 1);
    \coordinate[label={[anchor=east]$d$}] (D) at (0, 1);

    \draw[dashed] (B) -- (D);

    \foreach \n in {A, B, C, D}
      \filldraw (\n) circle (0.05cm);

    \draw (A) -- (B) -- (C) -- (D) -- (A);

  \end{tikzpicture}
  \caption{%
    If we use a cubical complex~(left; only a single square cell is
    shown) we do not have to choose an interpolation scheme for
    voxel-based data. Function values can be stored in the
    vertices~($a$, $b$, $c$, $d$) and interpolation happens along the
    edges.  For simplicial complexes, however, we need to convert the
    square into a triangle~(the same issue occurs in higher dimensions
    with cubes and tetrahedra, respectively). This conversion to
    triangles leaves us with two ways of interpolating that will
    typically lead to different results. In one case, we are
    interpolating between~$a$ and~$c$, in the other case between~$b$ and~$d$.
    Neither one of these edges exist in the original data, though.
  }
  \label{fig:Simplicial complex interpolation}
\end{figure}
%%%%%%%%%%%%%%%%%%%%%%%%%%%%%%%%%%%%%%%%%%%%%%%%%%%%%%%%%%%%%%%%%%%%%%%%

\autoref{fig:Simplicial complex interpolation} depicts the differences
between cubical complexes and simplicial complexes. The cubical
complex is `aligned' with a regular grid and does \emph{not} force us to
choose between an interpolation scheme. For a simplicial complex,
however, the calculation of topological features in dimensions~$1$
and~$2$ necessitates the creation of \mbox{$2$-simplices}, i.e.\
triangles. This, in turn, requires us to `pick' between two
triangulation schemes that result in different connectivities between
the original vertices. In the worst case, this could lead to subtle
differences in filtrations, since the new edges need to be weighted
accordingly.

%%%%%%%%%%%%%%%%%%%%%%%%%%%%%%%%%%%%%%%%%%%%%%%%%%%%%%%%%%%%%%%%%%%%%%%%
\subsection{fMRI pre-processing}\label{sec:fMRI pre-processing}
%%%%%%%%%%%%%%%%%%%%%%%%%%%%%%%%%%%%%%%%%%%%%%%%%%%%%%%%%%%%%%%%%%%%%%%%

The fMRI data acquisition used the following parameters: gradient-echo
EPI sequence: TR = \SI{2}{\second}, TE = \SI{30}{\milli\second}, flip
angle = \SI{90}{\degree}, matrix = $64 \times 64$, slices = 32,
and interleaved slice acquisition.
Data were collected using the standard Siemens \mbox{32-channel} head
coil for adults and older children.  One of two custom 32-channel
phased-array head coils was used for younger children~(smallest coil:
N~=~3; M~=~3.91, SD~=~0.42 years old; smaller coil: N~=~28; M~=~4.07,
SD~=~0.42, years old).
Acquisition parameters differed slightly across participants but all
fMRI data were re-sampled to have the same voxel size, namely
\SI{3}{\milli\meter} isotropic with
10\% slice gap.  A \mbox{T1-weighted} structural image was also collected for
all subjects~(MPRAGE sequence: GRAPPA = 3, slices = 176, resolution
= \SI{1}{\milli\meter} isotropic, adult coil FOV
= \SI{256}{\milli\meter}, child coils FOV = \SI{192}{\milli\meter}).
Imaging data were pre-processed using \texttt{fMRIPrep}
v1.1.8~\citep{esteban_fmriprep:_2019}. 

%%%%%%%%%%%%%%%%%%%%%%%%%%%%%%%%%%%%%%%%%%%%%%%%%%%%%%%%%%%%%%%%%%%%%%%%
\subsection{Baselines}\label{sec:Baselines}
%%%%%%%%%%%%%%%%%%%%%%%%%%%%%%%%%%%%%%%%%%%%%%%%%%%%%%%%%%%%%%%%%%%%%%%%

As additional comparison partners, we calculate a time point
correlation matrix and a spatial correlation matrix~(see
\autoref{sec:Related work}). These matrices are calculated from the
time-varying fMRI data of a single participant, which is a 4D~tensor
indexed by time steps and spatial coordinates. By `unravelling'
the spatial dimensions of the tensor~(using a \emph{row-major ordering},
for example), the 4D~tensor becomes a 2D~tensor, i.e.\ a matrix in which
each row corresponds to a single time step, and the columns correspond
to voxels in the aforementioned order. From this $m \times N$ matrix,
where $m$ denotes the number of time steps as in the main paper and $N$
denotes the total number of voxels, we can calculate Pearson product-moment correlation coefficients.
If we do this for the original matrix, we obtain an $m \times m$ \emph{time
point correlation matrix}, which we denote by \textsc{baseline-tt}~(i.e.\ a time-by-time matrix).

%%%%%%%%%%%%%%%%%%%%%%%%%%%%%%%%%%%%%%%%%%%%%%%%%%%%%%%%%%%%%%%%%%%%%%%%
\begin{figure}[tbp]
  \centering
  \includegraphics[width=0.30\linewidth]{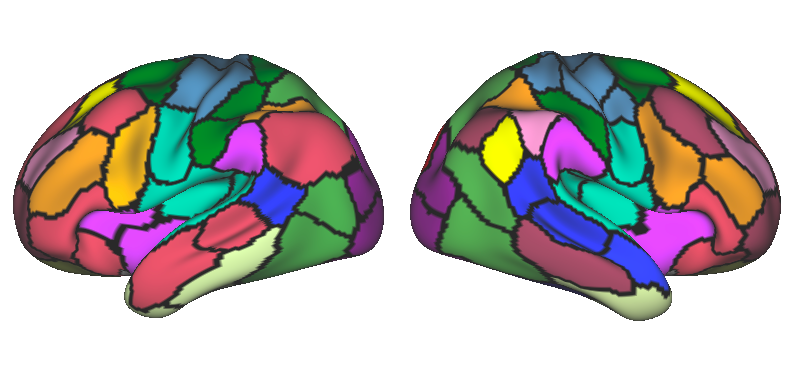}
  \quad
  \includegraphics[width=0.30\linewidth]{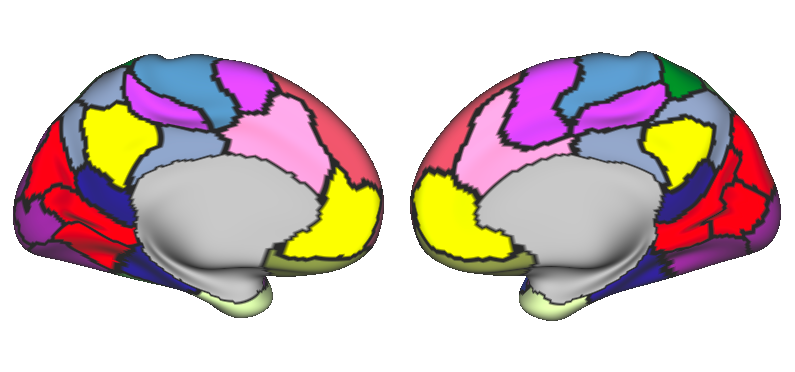}
  \caption{%
    A schematic illustration of the parcels used to make the 
    computation of a full correlation matrix computationally
    feasible. This only pertains to the \textsc{baseline-pp}
    method.
  }
  \label{fig:Parcellated data}
\end{figure}
%%%%%%%%%%%%%%%%%%%%%%%%%%%%%%%%%%%%%%%%%%%%%%%%%%%%%%%%%%%%%%%%%%%%%%%%

Conversely, we can transpose the matrix to obtain a voxel-by-voxel
correlation matrix---referred to as a \emph{full correlation
matrix}~(FCMA). This matrix has dimensions $N \times N$, though, which
is computationally prohibitive for most applications.
As a more feasible calculation, we
calculate the spatial correlation matrix from a parcellated data set. We use $100$ parcels from 17 functional networks \citet{Schaefer17},
depicted in \autoref{fig:Parcellated data}, so that we obtain a $100
\times 100$ correlation matrix, which we refer to as
\textsc{baseline-pp} to indicate that parcellated data was used to
obtain this matrix.

As additional comparison partners, we follow a more conventional
topological data analysis pipeline and calculate persistence images from
a set of correlation matrices. The matrix is treated as the adjacency
matrix of a fully-connected graph, and we use a filtration that is
specifically geared towards the analysis of such `correlation
graphs'~\citep{Chung17}. Following our own pipeline, we convert the
resulting persistence diagrams into persistence images~(using smoothing
values $\sigma \in \{0.1, 1.0\}$ and resolutions $r \in \{10, 20\}$,
respectively), and report the best performance for the age prediction
task. We denote the corresponding methods by \textsc{tt-corr-tda} and
\textsc{pp-corr-tda}, depending on which correlation matrix was used for
the calculation.

Last, as an ablation study, we use parcellated data with the same
parcels as above and assign the respective values to the original
cubical volume; each voxel corresponding to the same parcel is assigned
the same value. This has the effect of coarsening information but also
removing noise; while not decreasing the number of voxels in the data,
it will decrease the number of topological features that have to be
considered. We mark the results obtained using this technique with
\textsc{parcellated}.

%%%%%%%%%%%%%%%%%%%%%%%%%%%%%%%%%%%%%%%%%%%%%%%%%%%%%%%%%%%%%%%%%%%%%%%%
\subsection{Age prediction experimental details}\label{sec:Age prediction}
%%%%%%%%%%%%%%%%%%%%%%%%%%%%%%%%%%%%%%%%%%%%%%%%%%%%%%%%%%%%%%%%%%%%%%%%

For the age prediction experiment from \autoref{sec:Static analysis}, we
use a ridge regression classifier with internal leave-one-out
cross-validation~\citep{scikit-learn} for its regularisation strength
parameter $C \in \{0.1, 1.0, 10.0\}$. Internally, the classifier
optimises the $R^2$ score, i.e.\ the coefficient of determination. We
report the correlation coefficient in the table in order to be aligned
with the reporting in previous publications, though. We standardise
\emph{all} features prior to using them for the classifier. The
classifier is then used in a leave-one-out cross-validation scheme.

%%%%%%%%%%%%%%%%%%%%%%%%%%%%%%%%%%%%%%%%%%%%%%%%%%%%%%%%%%%%%%%%%%%%%%%%
\begin{table}[bt]
  \caption{%
    Additional experimental results for the age prediction tasks. In
    contrast to the table in the main paper, here we show both the
    correlation coefficient~(CC; higher values are preferable~$\uparrow$) and the
    mean squared error~(MSE; lower values are preferable) whenever available.
  }
  \vspace{0.50\baselineskip}
  \label{tab:Age prediction extended}
  \robustify\bfseries
  \centering
  \begin{tabular}{lSSSSSS}
    \toprule
    Method & \multicolumn{2}{c}{BM} & \multicolumn{2}{c}{OM} & \multicolumn{2}{c}{XM}\\
    \midrule
                                      &           {CC $\uparrow$} &          {MSE $\downarrow$} &           {CC $\uparrow$} &          {MSE $\downarrow$} &           {CC $\uparrow$} &          {MSE $\downarrow$}\\
    \midrule
    \textsc{baseline-tt}              &           0.09 &         10.15  &           0.02 &          13.81 &           0.24 &           7.19\\
    \textsc{baseline-pp}              &           0.41 &          6.23  &           0.40 &           6.40 &           0.40 &           6.65\\
    \textsc{tt-corr-tda}              &           0.17 &          10.04 &           0.11 &          12.57 &           0.23 &           9.76\\
    \textsc{pp-corr-tda}              &           0.25 &          10.34 &           0.27 &           9.68 &           0.23 &           9.94\\
    \midrule
    \textsc{srm}                      &           0.44 &          6.05  &          {---} &          {---} &          {---} &          {---}\\
    \midrule
    $\|\diagram\|_1$                  &           0.46 &           4.27 &           0.67 &           2.95 &           0.48 &           4.17\\
    $\|\diagram\|_1$ parcellated      &           0.32 &           4.91 &           0.50 &           4.06 &           0.34 &           4.76\\
    $\|\diagram\|_\infty$             &           0.61 &           3.38 & \bfseries 0.77 & \bfseries 2.20 & \bfseries 0.73 & \bfseries 2.53\\
    $\|\diagram\|_\infty$ parcellated & \bfseries 0.67 & \bfseries 2.99 &           0.50 &           4.04 &           0.33 &           4.81\\
    \bottomrule
  \end{tabular}
\end{table}
%%%%%%%%%%%%%%%%%%%%%%%%%%%%%%%%%%%%%%%%%%%%%%%%%%%%%%%%%%%%%%%%%%%%%%%%

Since the input features have different cardinalities~(the parcellated
voxel-by-voxel matrix, for example, has a cardinality of $100^2$, which
will lead to severe overfitting, we reduce the baseline matrices to
$100$ dimensions using principal component analysis. Moreover, to
demonstrate the impact of our topological summary statistics, we only
use summary statistics from the second half of the time series,
resulting in \emph{less} than $100$ features. We observe that the
results are highly stable; even reducing the number of selected features
to less than $10$ has no noticeable effect on the resulting regression
model, indicating the informativeness of topology for this task.

\autoref{tab:Age prediction extended} shows extended results for this
experiment, including MSE values~(another goodness-of-fit measure) that
were excluded from the table in the main paper because the
incomparability to existing methods. To reproduce the values in this
table, please use the provided \texttt{predict\_age.py} script.

%%%%%%%%%%%%%%%%%%%%%%%%%%%%%%%%%%%%%%%%%%%%%%%%%%%%%%%%%%%%%%%%%%%%%%%%
\subsection{Proof of the stability theorem}\label{sec:Stability proof}
%%%%%%%%%%%%%%%%%%%%%%%%%%%%%%%%%%%%%%%%%%%%%%%%%%%%%%%%%%%%%%%%%%%%%%%%

\begin{proof}
  Let $\volume := [0,1]^3$; this is not a restriction because
  fMRI volumes are bounded, so they are always homeomorphic to this
  `standard cube'. Hence, $\volume$ is a compact metric space that can
  be triangulated. Since~$f$ and~$g$ are continuous functions~(at least
  this is the `idealised' view in which we have access to an infinite number of
  samples), the stability follows from the main theorem of \citet{Cohen-Steiner07}.
\end{proof}

%%%%%%%%%%%%%%%%%%%%%%%%%%%%%%%%%%%%%%%%%%%%%%%%%%%%%%%%%%%%%%%%%%%%%%%%
\subsection{Across-cohort variability analysis}\label{sec:Across-cohort variability analysis}
%%%%%%%%%%%%%%%%%%%%%%%%%%%%%%%%%%%%%%%%%%%%%%%%%%%%%%%%%%%%%%%%%%%%%%%%

%%%%%%%%%%%%%%%%%%%%%%%%%%%%%%%%%%%%%%%%%%%%%%%%%%%%%%%%%%%%%%%%%%%%%%%%
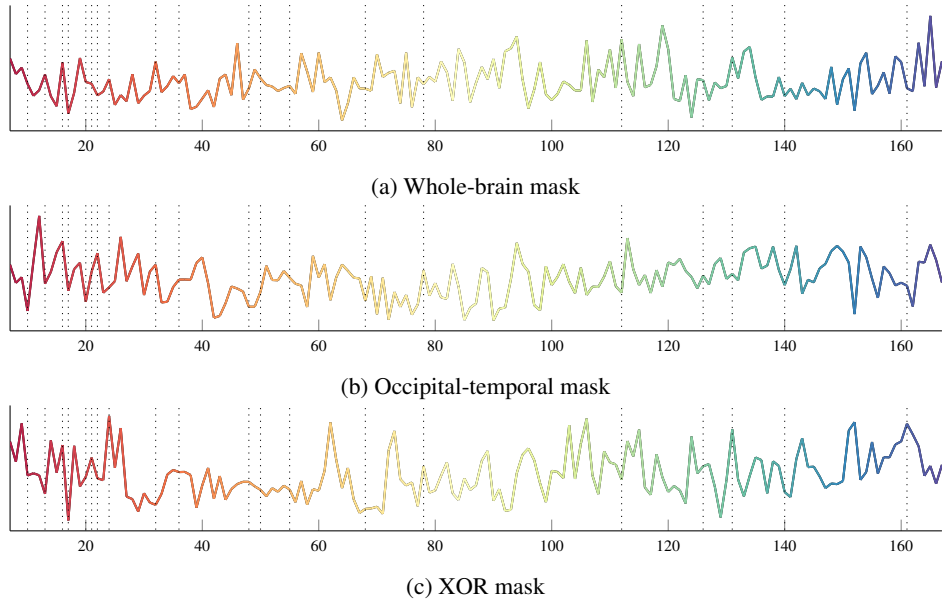
\begin{figure}[tbp]
  \centering
  \subcaptionbox{Whole-brain mask}{%
    \begin{tikzpicture}
      \begin{axis}[
        axis x line*     = bottom,
        axis y line*     = left,
        width            = \linewidth,
        height           = 3.25cm,
        enlarge x limits = false,
        point meta min   = 0,
        point meta max   = 167,
        ymajorticks      = false,
        ticklabel style  = {font = \tiny},
        label style      = {font = \tiny},
        colormap/Spectral,
        % Do *not* convert this to math mode
        tick label style = {
          /pgf/number format/assume math mode = true
        },
      ]
        \addplot[thick, mesh, black, no marks] table[x = time, y = mean, col sep = comma]{Data/across_cohort_variability/brainmask_sigma1.0_r20.csv};
        \addplot[thick, mesh, scatter, scatter src = x, no marks] table[x = time, y = mean, col sep = comma]{Data/across_cohort_variability/brainmask_sigma1.0_r20.csv};

        \pgfplotsinvokeforeach{10, 13, 16, 17, 20, 21, 22, 24, 32, 36, 48, 50, 55, 68, 78, 112, 126, 131, 140, 161}
        {
          \draw[dotted] (#1, 0) -- (#1, 1);
        }
      \end{axis}
    \end{tikzpicture}
  }
  \subcaptionbox{Occipital-temporal mask}{%
    \begin{tikzpicture}
      \begin{axis}[
        axis x line*     = bottom,
        axis y line*     = left,
        width            = \linewidth,
        height           = 3.25cm,
        enlarge x limits = false,
        point meta min   = 0,
        point meta max   = 167,
        ymajorticks      = false,
        ticklabel style  = {font = \tiny},
        label style      = {font = \tiny},
        colormap/Spectral,
        %
        % Do *not* convert this to math mode
        tick label style = {
          /pgf/number format/assume math mode = true
        },
      ]
        \addplot[thick, mesh, black, no marks] table[x = time, y = mean, col sep = comma]{Data/across_cohort_variability/occipitalmask_sigma1.0_r20.csv};
        \addplot[thick, mesh, scatter, scatter src = x, no marks] table[x = time, y = mean, col sep = comma]{Data/across_cohort_variability/occipitalmask_sigma1.0_r20.csv};

        \pgfplotsinvokeforeach{10, 13, 16, 17, 20, 21, 22, 24, 32, 36, 48, 50, 55, 68, 78, 112, 126, 131, 140, 161}
        {
          \draw[dotted] (#1, 0) -- (#1, 1);
        }
      \end{axis}
    \end{tikzpicture}
  }
  \subcaptionbox{XOR mask}{%
    \begin{tikzpicture}
      \begin{axis}[
        axis x line*     = bottom,
        axis y line*     = left,
        width            = \linewidth,
        height           = 3.25cm,
        enlarge x limits = false,
        point meta min   = 0,
        point meta max   = 167,
        ymajorticks      = false,
        ticklabel style  = {font = \tiny},
        label style      = {font = \tiny},
        colormap/Spectral,
        %
        % Do *not* convert this to math mode
        tick label style = {
          /pgf/number format/assume math mode = true
        },
      ]
        \addplot[thick, mesh, black, no marks] table[x = time, y = mean, col sep = comma]{Data/across_cohort_variability/xormask_sigma1.0_r20.csv};
        \addplot[thick, mesh, scatter, scatter src = x, no marks] table[x = time, y = mean, col sep = comma]{Data/across_cohort_variability/xormask_sigma1.0_r20.csv};

        \pgfplotsinvokeforeach{10, 13, 16, 17, 20, 21, 22, 24, 32, 36, 48, 50, 55, 68, 78, 112, 126, 131, 140, 161}
        {
          \draw[dotted] (#1, 0) -- (#1, 1);
        }
      \end{axis}
    \end{tikzpicture}
  }
  \caption{%
    Across-cohort variability curves for the different masks. The dotted
    lines represent the events. Generally, events are aligned with local
    extrema of the curves.
  }
  \label{fig:Across-cohort variability curves}
\end{figure}
%%%%%%%%%%%%%%%%%%%%%%%%%%%%%%%%%%%%%%%%%%%%%%%%%%%%%%%%%%%%%%%%%%%%%%%%

For the across-cohort variability analysis, \autoref{fig:Across-cohort
variability curves} shows the `raw' curves for each of the masks,
annotated with the respective events.
As described in \autoref{sec:Variability analysis}, we \emph{pool}
variability for all events and analyse the resulting histograms. This
construction loses some information, but is a simple way to assess
overall differences in variability.

%%%%%%%%%%%%%%%%%%%%%%%%%%%%%%%%%%%%%%%%%%%%%%%%%%%%%%%%%%%%%%%%%%%%%%%%
\subsection{Curvature analysis}\label{sec:Curvature analysis}
%%%%%%%%%%%%%%%%%%%%%%%%%%%%%%%%%%%%%%%%%%%%%%%%%%%%%%%%%%%%%%%%%%%%%%%%

The curvature~$\kappa$ of a differentiable curve measures how sharply the
trajectory curves at a given point \emph{on} the curve. A circle, for
example, always has curvature of~$1$ at each point, while a straight
line, by contrast, always has a curvature of~$0$.
We hypothesise that the curvature of brain state trajectories can help
to further characterise subjects by looking at topological activity from
a geometric point of view.

\begin{figure}[t]
	\centering
	\includegraphics[width=0.65\linewidth]{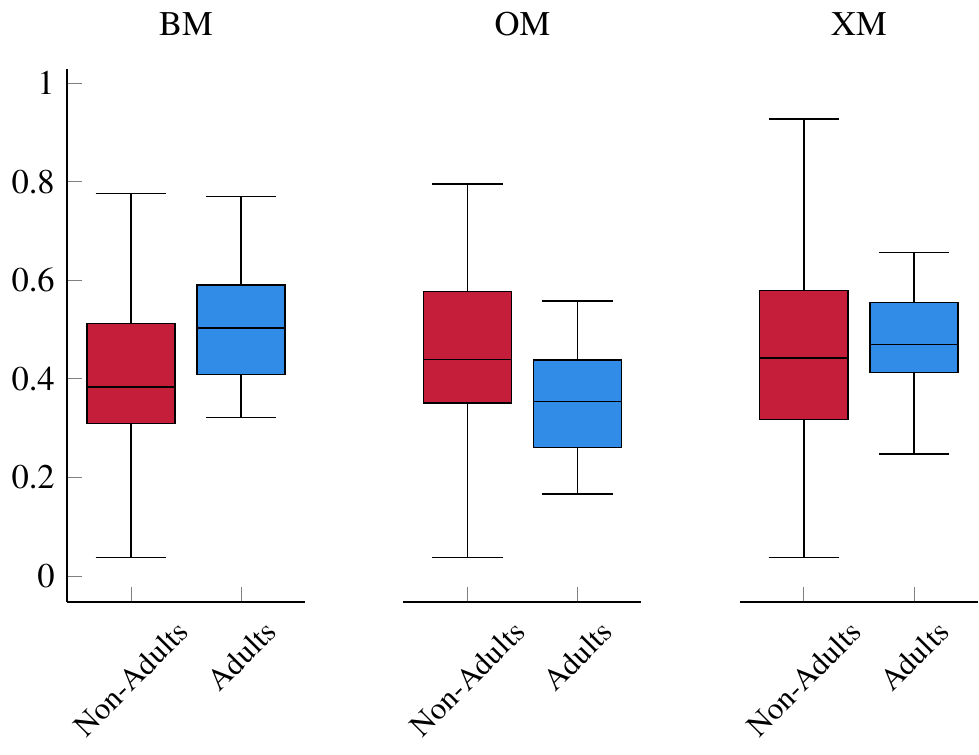}
	\caption{%
  Distribution of brain state trajectory curvature values at event
  boundaries. The distributions differ significantly for~BM~($p_\mathrm{KS}=0.00406$ 
  and $p_{t}=0.0327$) and~OM~($p_{\mathrm{KS}}=0.0276$ and $p_{t}=0.0402$), but
  not so for XM~($p_\mathrm{KS}=0.0861$ and $p_{t}=0.282$).
  }
	\label{fig:Curvature}
\end{figure}

Let $x(t)$ be the $x$-coordinate of a brain trajectory~(as shown in
\autoref{fig:Brain state trajectories}) at time~$t$ and $y(t)$ its
respective $y$-coordinate. Furthermore, let us define~$\dot{x}$ as the
first derivative of~$x$ with respect to~$t$; equivalently, $\ddot{x}$
denotes the second derivative.
Curvature can then be expressed as
\begin{equation}
  \kappa = \frac{|\dot{x}\ddot{y}-\dot{y}\ddot{x}|}{(\dot{x}^2 + \dot{y}^2)^{\frac{3}{2}}}
\end{equation}
Notice that curvature is an inherently \emph{local} quantity.
We computed $\kappa$ for all brain state trajectories of all cohorts and
for all three segmentation masks. We then investigated the differences
in~$\kappa$ around event boundaries, similar to the variability analysis
in \autoref{sec:Variability analysis}.
\autoref{fig:Curvature} shows the distribution of curvature values when
stratifying the subjects into \emph{adults} and \emph{non-adults}. We
find significant differences~(at the $\alpha = 0.05$ level) in the
distribution of values in terms of a two-sided Kolmogorov--Smirnov test~($p_\mathrm{KS}$)
and in terms of a $T$-test~($p_t$) for both~BM and OM,
whereas the differences in~XM are not considered to be significant.

%%%%%%%%%%%%%%%%%%%%%%%%%%%%%%%%%%%%%%%%%%%%%%%%%%%%%%%%%%%%%%%%%%%%%%%%
%%%%%%%%%%%%%%%%%%%%%%%%%%%%%%%%%%%%%%%%%%%%%%%%%%%%%%%%%%%%%%%%%%%%%%%%

\end{document}